\documentstyle[aaspp4,aabib]{article}
\input epsf
\newcommand{\Grad}{\bf{\nabla}}

\def\inlinefig{1}

\begin{document}
\title{Theoretical Models of Polarized Dust Emission from Protostellar Cores}

\author{Paolo Padoan,\footnote{ppadoan@cfa.harvard.edu}
Alyssa Goodman,
\affil{Harvard-Smithsonian Center for Astrophysics, Cambridge, MA 02138}}
\author{Bruce Draine,
\affil{Princeton University Observatory, Peyton Hall, Princeton, NJ 08544}} 
\author{Mika Juvela, 
\affil{Helsinki University Observatory, T\"ahtitorninm\"aki, P.O.Box 14,
SF-00014 University of Helsinki, Finland}}
\author{\AA ke Nordlund}
\affil{Copenhagen Astronomical Observatory, and Theoretical Astrophysics Center, 
2100 Copenhagen, Denmark}
\author{\"{O}rn\'{o}lfur Einar R\"{o}gnvaldsson}
\affil{Nordic Institute for Theoretical Physics, 2100 Copenhagen, Denmark }

\begin{abstract}

We model the polarized thermal dust emission from protostellar cores that are
assembled by super--sonic turbulent flows in molecular clouds.

Self--gravitating cores are selected from a three dimensional simulation of
super--sonic and super--Alfv\'{e}nic magneto--hydrodynamic (MHD) turbulence. 
The polarization is computed in two ways. In model $A$ it is assumed that
dust properties and grain alignment efficiency are uniform; in 
model $B$ it is assumed that grains are not aligned at visual extinction 
larger than $A_{V,0}=3$~mag, consistent with theoretical expectations for
grain alignment mechanisms. Instead of using a specific set of grain
properties, we adopt a maximum degree of polarization $P_{max}=15$~\%. 
Results are therefore sensitive mainly to the topology of the magnetic field
(model $A$) and to the gas distribution that determines the distribution of 
$A_V$ (model $B$). Furthermore, the radiative transfer in the MHD model is 
solved with a non--LTE Monte Carlo method, to compute spectral maps
of the J=1--0 transition of CS. The CS spectral maps are used to 
estimate the turbulent velocity, as in the observations.

The main results of this work are: i) Values of $P$ between 1 and 10\% (up to
almost $P_{max}$) are typical, despite the super--Alfv\'{e}nic nature of the
turbulence; ii) A steep decrease of $P$ with increasing values of 
the sub-mm dust continuum intensity $I$ is always found in self--gravitating
cores selected from the MHD simulations, if grains are not aligned above a 
certain value of visual extinction $A_{V,0}$ (model $B$); iii) The same behavior
is hard to reproduce if grains are aligned independently of $A_{V}$ (model $A$);
iv) The Chandrasekhar--Fermi formula, corrected by a factor $f\approx 0.4$,
provides an approximate estimate of the average magnetic field strength
in the cores. 

Sub--mm dust continuum polarization maps of quiescent protostellar 
cores and Bok globules have recently been obtained. They always
show a decrease in $P$ with increasing value of $I$ consistent with
the predictions of our model $B$. We therefore conclude that sub--mm 
polarization maps of quiescent cores do not map the magnetic field 
inside the cores at visual extinction larger than $A_{V,0}\approx 3$~mag. 
The use of such maps to constrain models of protostellar core 
formation and evolution is questionable. This conclusion suggests 
that there is no inconsistency between the results
from optical and near--IR polarized absorption of background stars,
and the observed polarization of sub-mm dust continuum from
quiescent cores. In both cases, grains at large visual extinction
appear to be virtually unaligned.

\end{abstract}


\section{Introduction}

The topology of the magnetic field inside and around protostellar cores can be 
predicted with models for their formation and evolution. Models of sub--critical
cores contracting under the effect of gravity and ambipolar drift, for example,
predict a rather uniform magnetic field roughly perpendicular to the core major
axis. Such models can be tested if the magnetic field topology is constrained 
observationally.

Polarization maps of sub--mm thermal dust emission have recently been obtained 
for a number of protostellar cores and Bok globules (Minchin, Bonifacio \&
Murray 1996; Glenn, Walker \& Young 1999; Greaves et al. 1999; Coppin et al. 2000; 
Davis et al. 2000; Matthews \& Wilson 2000; Vall\'{e}e, Bastien \& Greaves 2000;  
Ward--Thompson et al. 2000; Henning et al. 2001). Results are still sparse, 
but are readily used to try to test theoretical 
models of core formation and evolution (Glenn, Walker \& Young 1999; Coppin et al. 
2000; Davis et al 2000; Ward--Thompson et al. 2000). Since the interpretation of the 
observations is hardly unique, it is useful to compute polarization maps of the theoretical 
models, where the full three--dimensional information is available, and to infer their 
observable properties. Theoretical polarization maps have been previously computed by 
Wardle \& K\"{o}nigl (1990) for the molecular disk at the Galactic center and by
Fiege \& Pudritz (2000) for molecular cloud filaments.

\nocite{Minchin+96} \nocite{Glenn+99} \nocite{Greavens+99}
\nocite{Davis+2000} \nocite{Ward--Thompson+2000} \nocite{Vallee+2000}
\nocite{Henning+2001} \nocite{Coppin+2000} \nocite{Matthews+Wilson2000}

\nocite{Fiege+Pudritz2000} \nocite{Wardle+Konigl1990}

In this work we assume that protostellar cores are assembled by super--sonic
turbulent flows in molecular clouds. We have shown in a previous work that the
formation of cores by turbulent shocks provides an excellent interpretation
of the relation between integrated intensity and rms velocity in molecular
clouds (Padoan et al. 2001). Alternative interpretations of this newly discovered
property of molecular clouds have not been proposed so far. Furthermore, 
super--sonic turbulence is ubiquitously observed in molecular clouds, and 
therefore models of core formation and evolution must take into account
this turbulent environment self--consistently. We do so by selecting
self--gravitating cores from a numerical simulation of super--sonic,
super--Alfv\'{e}nic, self--gravitating magneto--hydrodynamic (MHD)
turbulence. The simulation is intended to represent a molecular cloud region
with size $L=6.25$~pc, rms turbulent velocity $\sigma_v=3.0$~km/s
and average gas density $\langle n \rangle=320$~cm$^{-3}$. These are typical 
values for molecular clouds according to Larson's relations (Larson 1981). 
The three selected cores have a mass of $\approx 60$~M$_{\odot}$ and a maximum 
gas density $\sim 10^5$~cm$^{-3}$.

\nocite{Padoan+2001cores} \nocite{Larson81}

The polarized thermal dust emission is computed in two ways. In model A 
it is assumed that dust properties and grain alignment efficiency are 
uniform; in model B it is assumed that grains are not aligned at visual 
extinction larger than $A_{V,0}=3$~mag, in line with theoretical expectations
for grain alignment mechanisms (Lazarian, Goodman \& Meyers 1997). 
Instead of using a specific set 
of grain properties, we adopt a maximum degree of polarization $P_{max}=15$~\%. 
Results are therefore sensitive mainly to the topology of the magnetic field
(model A) and to the gas distribution that determines the distribution of 
$A_V$ (model B). A Monte Carlo method is used to compute the distribution
of $A_V$ in the MHD model.

We find that despite the relatively weak magnetic field averaged over the whole
computational box, $\langle B \rangle=3.3$~$\mu$G and 
$\langle B^2 \rangle^{1/2}=7.1$~$\mu$G,
the degree of polarization is typically between 1 and 10\% (as observed), 
spanning a range of values almost up to $P_{max}$. In protostellar cores 
(Davis et al. 2000) and Bok globules (Henning et al. 2000), $P$ is always found
to decrease with increasing sub-mm dust continuum intensity $I$. This 
is consistent with our MHD model only if grains are not aligned above a certain 
value of the visual extinction, $A_{V,0}$ (model B), in agreement with 
previous results from optical and near--IR polarized absorption of
background stars (Goodman et al. 1995; Gerakines et al. 1995; Arce et al. 1998).

\nocite{Goodman+95} \nocite{Gerakines+95} \nocite{Arce+98}

We also solve the radiative transfer in the MHD model with a non--LTE 
Monte Carlo method, 
to compute spectral maps of the J=1--0 transition of CS.
We use these maps to estimate the turbulent velocity dispersion in the same way as
in the observations. This turbulent velocity dispersion is used, together with the
dispersion in the polarization angle, to infer the magnetic field strength
using the Chandrasekhar--Fermi formula (Chandrasekhar \& Fermi 1953). 
It is found that the volume--averaged three dimensional magnetic field strength 
can be estimated, despite the large dispersion in the polarization angle, if 
the Chandrasekhar--Fermi formula is used with a coefficient $f\approx 0.4$.

\nocite{Chandrasekhar+Fermi53}

The polarization properties of turbulent molecular clouds depend on the average
magnetic field strength, relative to the kinetic energy of turbulence, as
measured by the rms Alfv\'{e}nic Mach number, ${\cal M}_a$, that is the ratio of 
the rms flow velocity and the Alfv\'{e}n velocity, (Clemens et al. 2000; 
Ostriker, Stone \& Gammie 2001). A complete statistical study of the polarization 
properties of turbulent clouds with different values of ${\cal M}_a$ will be 
presented separately, and will be more useful in the future, when a larger number
of polarization maps of sub--mm and far--IR dust emission will be available (especially
if future satellite missions of the type of M4 (Clemens et al. 2000)
will be dedicated to this purpose). In the present work we focus on the 
properties of a few cores, to provide insight in the interpretation of recent 
sub--mm observations.

\nocite{Clemens+2000} \nocite{Ostriker+2001}

The MHD models are presented in the next section, while in \S~3 we describe the method
for computing the polarized dust emission. The selection of self--gravitating cores
from the synthetic continuum map is discussed in \S~4 and the observational results
are summarized in \S~5. In \S~6 the Chandrasekhar--Fermi
formula is tested on the selected cores. Results are discussed in \S~7, and 
conclusions are summarized in \S~8.

\section{The MHD Model}

We solve the compressible MHD equations:

\def\vv{{\bf v}}
\def\jj{{\bf j}}
\def\bb{{\bf B}}
\def\lnr{\ln\rho}
\def\div{\nabla\cdot}

\begin{equation}
\label{0}
{\partial \ln\rho \over \partial t} + \vv \cdot \nabla\lnr = - \div \vv,
\end{equation}

  \begin{equation}
   {\partial{\vv} \over \partial t}
   + {\vv\cdot\nabla\vv}
  =
   - {P\over\rho} \nabla \ln P
   + {1\over\rho} {\jj} \times {\bb} - {\Grad}{{\Phi}}
   + {\bf f},
  \label{1}
  \end{equation}

\begin{equation}
\label{4}
P=\rho \, T,
\end{equation}

\begin{equation}
{\partial{\bb} \over \partial t} = \nabla\times(\vv\times\bb),
\label{2}
\end{equation}

\begin{equation}
\jj = \nabla\times\bb,
\label{3}
\end{equation}

\begin{equation}
\nabla^2 \Phi = C \rho
\end{equation}

\noindent
plus numerical diffusion terms, and
with periodic boundary conditions. $\vv$ is the velocity, $\bb$ the
magnetic field, $\Phi$ the gravitational potential, ${\bf f}$ an external 
random force, $T$ the gas temperature and $T =$~const is assumed. 
The isothermal approximation is appropriate for the short cooling
time in relatively dense molecular gas (see discussion
in Padoan, Zweibel \& Nordlund 2000). The constant $C$ is given by:

\nocite{Padoan+2000ad}

\begin{equation}
C = \frac{4 \pi G l_0^2 \rho_0}{v_0^2},
\end{equation}
where the velocity is measured in units of $v_0$, length in units of $l_0$, 
time in units of $l_0/v_0$, and density in units of $\rho_0$.

For the purpose of this paper we have run a numerical simulation
of super--sonic, super--Alfv\'{e}nic and self--gravitating MHD 
turbulence, by solving numerically the equations above, in a $128^3$ mesh.
The initial density and magnetic fields
are uniform; the initial velocity is random, generated in
Fourier space with power only on the large scale. We also apply
an external random force, to drive the turbulence at a roughly constant
rms Mach number of the flow. This force is generated in Fourier
space, with power only on small wave
numbers ($1<k<2$), as the initial velocity.

We let the flow evolve for one dynamical time, defined as 
$t_{dyn}=L_0/\sigma_v$, where $L$ is the linear size of the 
computational box, and $\sigma_v$ is the rms flow velocity. 
A self--gravitating flow is never statistically relaxed. 
First a number of collapsing and accreting cores is generated, 
and later the cores begin to merge. The numerical
resolution (128$^3$ numerical mesh) allows only the description
of the initial phase of the collapse of single cores. Results
are progressively inaccurate at later times, when the numerical
resolution cannot cope with the exceedingly high density. We therefore 
interrupt the simulation at a time when most cores are just
recently formed and start to collapse, which is about one dynamical
time of the large scale.

Periodic boundary conditions and large scale external forcing
are justified by the fact that we simulate a region
of turbulent flow inside a larger turbulent molecular 
cloud. The rms Mach number of the flow is ${\cal M}_s\approx 11$, 
which corresponds to a linear size $L_0\approx 6.25$~pc, 
and an average gas density $<n>\approx 320$~cm$^{-3}$, 
using empirical Larson type relations (Larson 1981). 
The average magnetic field in this model is rather weak, as
justified by our previous work (Padoan \& Nordlund 1997, 1999), and 
such that the average magnetic energy is approximately
twice the thermal energy. Assuming a kinetic temperature $T=10$~K,
the rms flow velocity is $\sigma_v = 3.0$~km/s; the
average magnetic field strength is $\langle B\rangle = 3.3$~$\mu$G,
while $\langle B^2\rangle^{1/2} = 7.1$~$\mu$G .
Despite this low value of $\langle B\rangle$, strongly magnetized cores are
formed by the process of turbulent fragmentation (the maximum field strength 
in the simulation is 127.8~$\mu$G), due to compressions and stretching in the 
turbulent flow.
  
\nocite{Padoan+Nordlund97MHD}  \nocite{Padoan+Nordlund98MHD}

\ifnum\inlinefig=1
\begin{figure}[!th]
%\centerline
{\epsfxsize=8.2cm \epsfbox{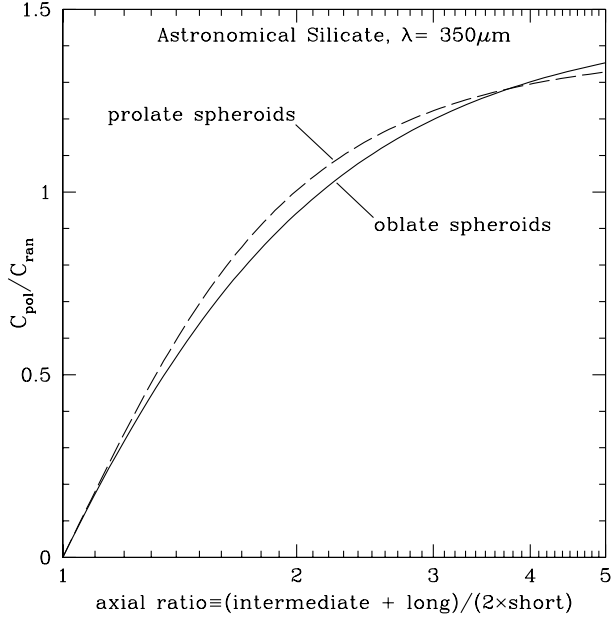}}
\caption[]{%
Grain cross section ratio as a function of grain axial ratio for
prolate spheroids (dashed line) and oblate spheroids (solid line),
at wavelength $\lambda=350$~$\mu$m. The grains are the
``astronomical silicate'' discussed by Draine \& Lee (1984), as 
modified by Li \& Draine (2001).}
\label{fig1}
\end{figure}
\fi

\section{Polarized Thermal Dust Emission}

We compute the polarized thermal dust emission from the MHD model following 
the formalism in Fiege \& Pudritz (2000) (see also Wardle \& K\"{o}nigl 1990). 
As in Fiege \& Pudritz (2000) we are interested in the thermal dust emission
at sub-mm wavelengths, and therefore the effect of self--absorption 
(Hildebrand et al. 1999) and scattering (Novak et al. 1989)
can be neglected. We further assume that the grain properties are 
constant\footnote{Fiege and Pudritz (2000) consider explicitly the 
contributions to the polarized emission from several grain species, which 
is not required here since we do not specify any particular set of grain
properties.} and the temperature is uniform. With these
assumptions, the Stokes parameters $Q$ and $U$ are proportional to the 
following integrals along the line of sight:
\begin{equation}
q=\int \rho \, \cos 2\psi \, \cos^2 \gamma \,\, ds,
\label{q}
\end{equation}
\begin{equation}
u=\int \rho \, \sin 2\psi \, \cos^2 \gamma \,\,ds,
\label{u}
\end{equation}
where $\rho$ is the gas density, $\psi$ the angle between the projection of $\bf B$
on the plane of the sky and the North, $\gamma$ the angle between the local $\bf B$
vector and the plane of the sky. The polarization angle $\chi$ is then given by:
\begin{equation}
\tan 2 \chi= \frac{u}{q}.
\label{tanchi}
\end{equation}

\nocite{Hildebrand+99}  \nocite{Novak+89}

The degree of polarization $P$, defined as 
\begin{equation}
P = \frac{\sqrt{Q^2+U^2}}{I},
\label{pdef}
\end{equation}
where $I$ is the intensity, is given by:
\begin{equation}
P = \alpha \, \frac{\sqrt{q^2+u^2}}{\Sigma-\alpha \Sigma_2},
\label{p}
\end{equation}
with
\begin{equation}
\Sigma = \int \rho \,\, ds,
\label{sigma}
\end{equation}
and
\begin{equation}
\Sigma_2 = \frac{1}{2}\int \rho \, (\cos^2 \gamma -2/3) \,\, ds,
\label{sigma2}
\end{equation}
The coefficient $\alpha$ in equation (\ref{p}) is defined as
\begin{equation}
\alpha = R \, F \, \frac{C_{pol}}{C_{ran}},
\label{alpha}
\end{equation}
where $R$, $F$, $C_{pol}$ and $C_{ran}$ are defined by Lee \& Draine (1985).
$C_{pol}$ is the grain polarization cross section, which differs for oblate 
spheroids,
\begin{equation}
C_{pol} = C_{\perp} - C_{\parallel},
\label{oblate}
\end{equation}
and prolate spheroids. 
\begin{equation}
C_{pol} = (C_{\parallel} - C_{\perp})/2,
\label{prolate}
\end{equation}
where $C_{\parallel}$ and $C_{\perp}$ are the polarization cross sections
for electric field parallel and perpendicular to the symmetry axis of the
grain.
$C_{ran}$ is the average cross section for randomly oriented grains,
\begin{equation}
C_{ran} = (2 C_{\perp} + C_{\parallel})/3.
\label{cran}
\end{equation}
$R$ is the Rayleigh polarization reduction factor due to imperfect grain
alignment, and $F$ the polarization reduction factor due to the turbulent
component of the magnetic field. In our case $F=1$ because the three dimensional
magnetic field is provided by the MHD model (assuming that no small scale
structure is unresolved in the numerical solution). However, in this work
we do not attempt to use specific models for the grain cross sections and for
the reduction factors, since our aim is to investigate the effect of the
magnetic field topology. We therefore adopt a constant value of $\alpha=0.15$, 
which is equivalent to assuming a reasonable value for the maximum degree of 
polarization, $P_{max}=15$\%, since
\begin{equation}
P_{max} = \frac{\alpha}{1-\alpha/6}.
\label{pmax}
\end{equation}

\nocite{Lee+Draine85}

\ifnum\inlinefig=1
\begin{figure}[!th]
\centerline
{\epsfxsize=15cm \epsfbox{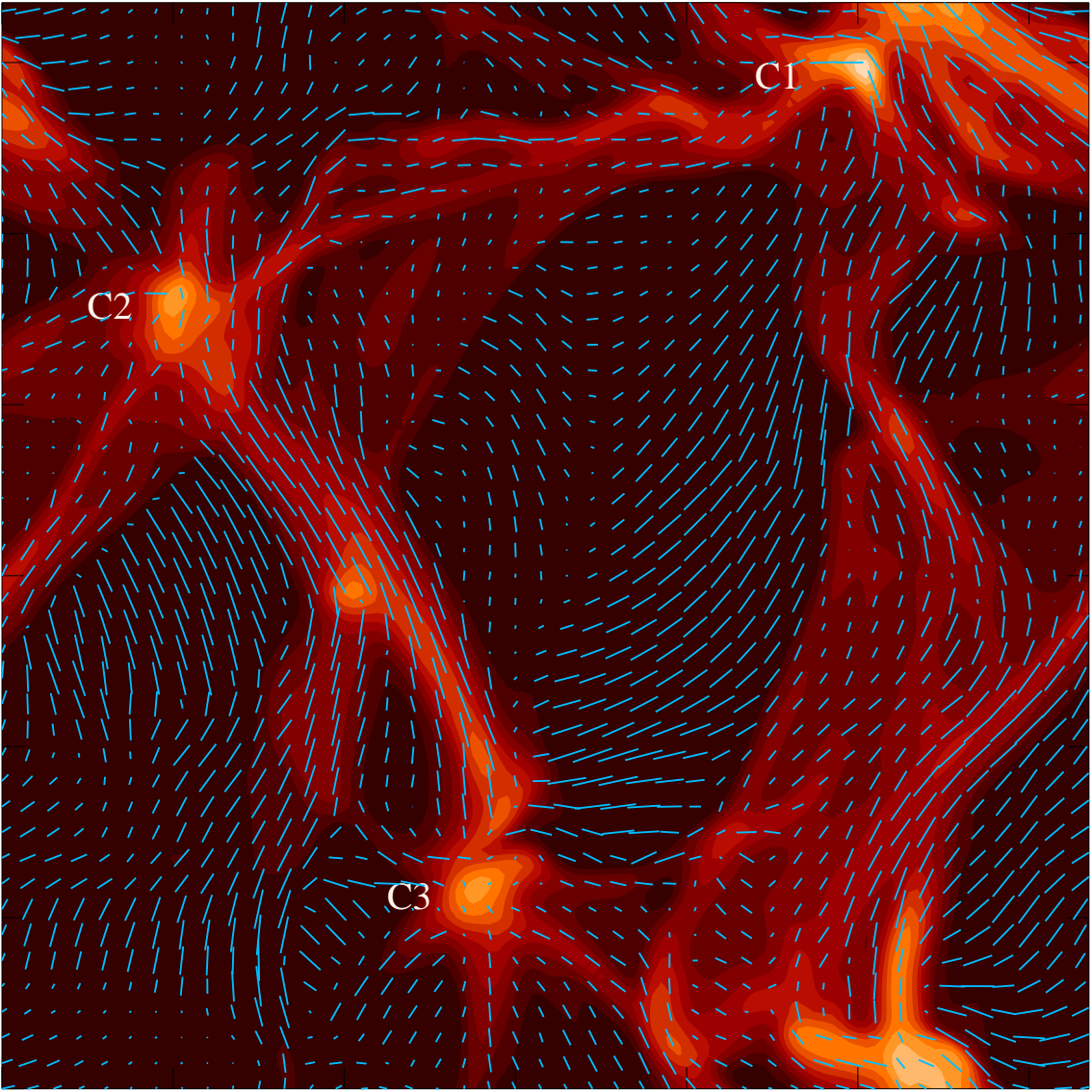}}
\caption[]{%
Polarization map from the MHD model (model $A$).
The length of the polarization vectors is proportional 
to the degree of polarization, with the longest vector corresponding to $P=14$~\%. 
Only one polarization vector every three computational cells is plotted. 
The contour map shows the sub--mm dust continuum intensity $I$.
The position of the self--gravitating cores C1, C2 and C3 (see text in \S4) 
is also shown.}
\label{fig2}
\end{figure}
\fi

\ifnum\inlinefig=1
\begin{figure}[!th]
\centerline
{\epsfxsize=15cm \epsfbox{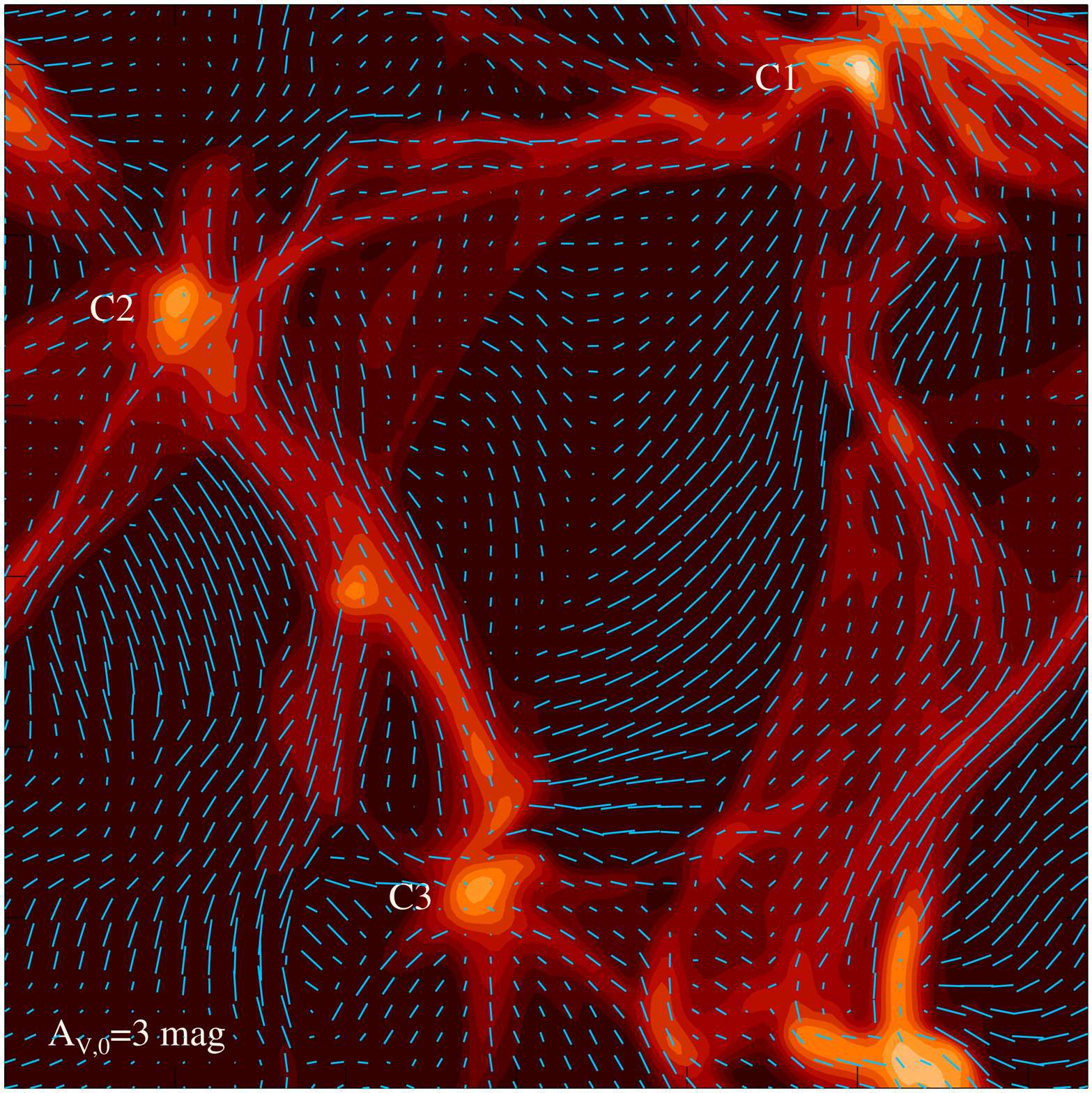}}
\caption[]{%
As in Figure~1, but assuming unaligned grains above $A_{V,0}=3$~mag (model $B$).}
\label{fig3}
\end{figure}
\fi

In Figure~1 we have plotted $C_{pol}/C_{ran}$ of astronomical silicate
spheroids, for ``perfect spinning alignment'', as a 
function of the ``axial ratio'', defined as  $(a+b)/(2c)$, where
$a$, $b$ and $c$ are the long, intermediate and short axis respectively.
For prolate spheroids $b=c$; for oblate spheroids $a=b$.
With this definition, the ratio $C_{pol}/C_{ran}$ is nearly the same, 
at given "axial ratio", for prolate and oblate spheroids.
The astronomical silicate is that discussed 
by Draine \& Lee (1984), as modified by Li \& Draine (2001).
For wavelengths $\lambda$ in the range $100 \,\mu{\rm m} < \lambda < 1 \,{\rm mm}$, 
$C_{pol}/C_{ran}$ is almost independent of $\lambda$. Figure~1 is computed for
$\lambda=350$~$\mu$m. The adopted value of $\alpha=0.15$ would require only moderate 
deviations from sphericity, provided the degree of alignment $R$ is reasonably 
high. If for example $R \approx 0.33$, a reasonable value of approximately 1.3 for
the axial ratio gives the assumed value of $\alpha$.

\nocite{Draine+Lee84}  \nocite{Li+Draine2001}

We have computed polarization maps from the MHD model in two cases. In the
first case, or model $A$, it is assumed that grains are aligned independently 
of their visual extinction, $A_V$; in the second case, or model $B$, it is assumed 
that grains are not aligned at visual extinction larger than $A_{V,0}=3$~mag.
The intensity of the radiation field was computed with a Monte Carlo
method. Photon packages were sent into the cloud from the background and
the scatterings and absorption processes were simulated. In each cell the
number of incoming photons was registered and the intensity relative to
the background was used to compute the effective values of $A_V$.
The dust opacity was calculated from the relation between $N$(H$_2$) and
$A_V$ as given by Bohlin, Savage \& Drake (1978). The albedo of the
grains was 0.5 and the asymmetry factor of the scattering $g$=0.6.

\nocite{Bohlin+78}

The  polarization maps are plotted in Figure~2 for both model $A$ and in 
Figure~3 for model $B$.
The average magnetic field (volume--average over the whole computational box)
is oriented in the vertical direction. The length of the polarization vectors 
is proportional to the degree of polarization, with the longest vector 
corresponding to $P=14$~\%. The contour map shows the sub--mm dust continuum
intensity $I$. Model $B$ shows systematically low values of the degree of
polarization $P$ in the brightest regions, because at $A_V>3$~mag the dust contributes
to the total flux, but not to the polarized flux ($P$ is defined as the ratio
of polarized and total flux). This does not happen in model $A$, with the exception
of one region (core C3).

There is no obvious correlation between $P$ and $I$ over the entire map,
as shown in the scatter plots of Figure~4. $P$ spans a range of values
0.3~\%$<P<14$~\%, with $\langle P\rangle=5.6$~\%. Such relatively large 
values of $P$, comparable to typically observed values, are usually
found in models of super--Alfv\'{e}nic turbulence. This is due to the fact
that each line of sight is normally dominated by one particular dense filament 
(or edge--on sheet), where the magnetic field is rather uniform because it is 
amplified in the directions perpendicular to the shock compression that 
generates that filament (or sheet). The power spectrum of turbulence 
contains most of the energy on the largest scales, which constrains 
the magnetic field tangling, in the sense that the number of correlation lengths
of the magnetic field can never be $\gg 1$ (see the discussion in \S~7).

\ifnum\inlinefig=1
\begin{figure}[!th]
\leavevmode
{\epsfxsize=8cm \epsfbox{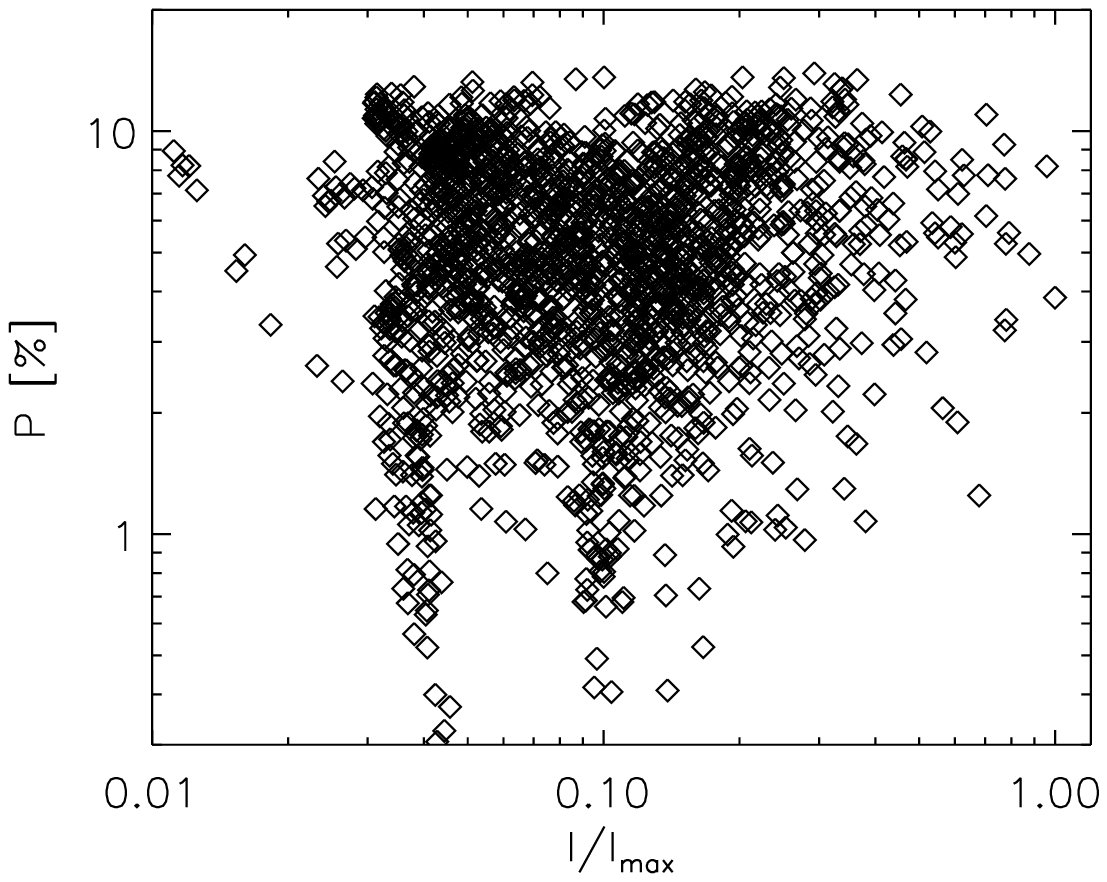}}
{\epsfxsize=8cm \epsfbox{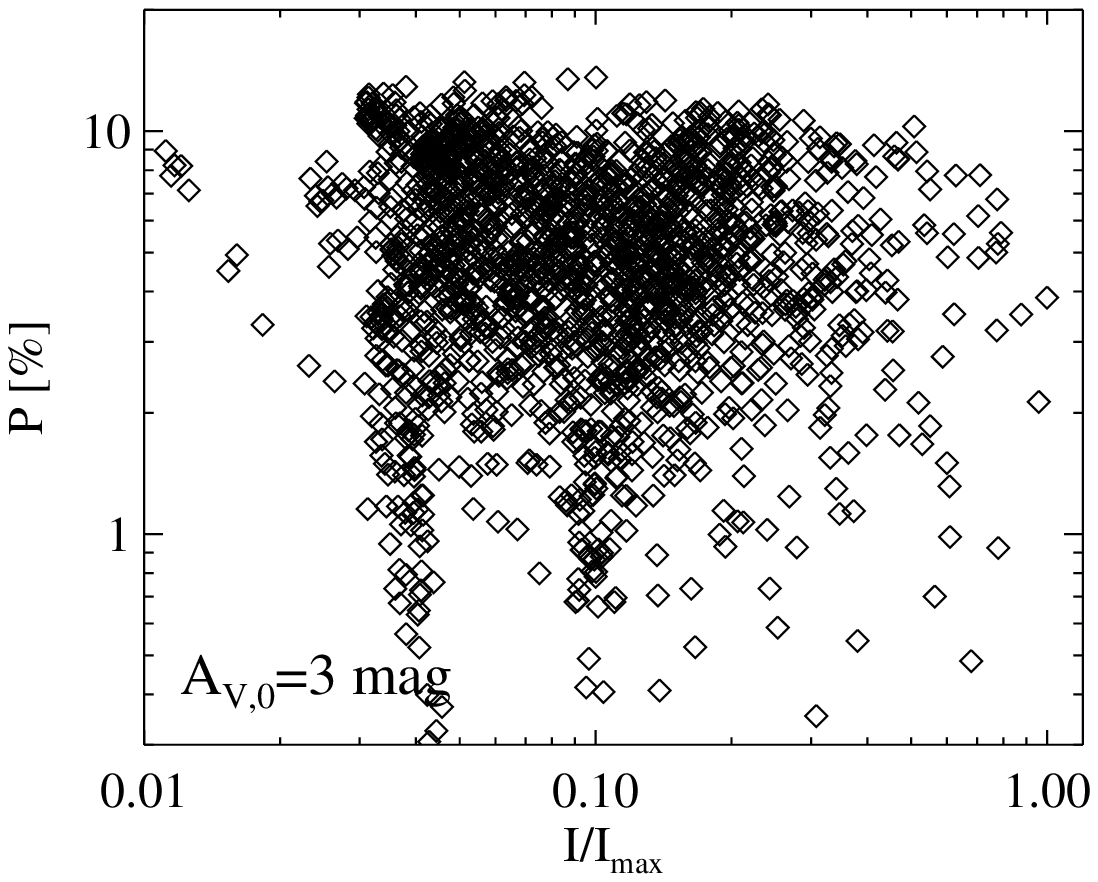}}
\caption[]{%
Left panel: Scatter plot of the degree of polarization versus the emission intensity
from the entire map in Figure~2. Only the polarization vectors shown in Figure~2
are used, that is one position every three computational cells. The average
degree of polarization is $\langle P\rangle=5.6$~\%. Right panel: As left panel,
but from the entire map in Figure~3, that is with unaligned grains above $A_{V,0}=3$~mag.}
\label{fig4}
\end{figure}
\fi

\section{Self--Gravitating Cores}

For our polarization study, we select  the brightest cores in the intensity 
map shown in Figure~2, in a similar way as protostellar cores would be 
selected from a sub--mm dust continuum map. Since we are interested in relating the 
observable properties of the cores with their intrinsic three dimensional 
structure, we have limited our selection to intensity maxima that are 
also well defined in the three dimensional MHD data--cube. We have excluded 
the remaining two or three bright cores that are not easily identified in 
the original three dimensional data--cube, because of the contribution of 
multiple structures along the line of sight.

We have selected three cores that we call C1, C2 and C3. Their position is shown
on the maps in Figures~2 and 3. We have verified, using the MHD data--cube, that 
the cores C1, C2 and C3 are gravitationally bound (gravitational energy in
excess of the kinetic energy of turbulence). While the core C1 is 
still being assembled by a turbulent shock (this is also indicated by its
position as a density peak in a strongly curved segment of a filament),
the core C3 is collapsing with a surrounding flow characterized
by accretion along numerous filaments converging towards the core.
The magnetic field around the core is roughly aligned with these accretion 
filaments. In the core C1 the magnetic 
field is roughly aligned with the filament where the core is located
(it is amplified in that direction by the shock compression), and is
therefore strongly curved inside the core. The core C2 seems to be 
an intermediate case between C1 and C3.

The collapsing/accreting configuration of the core C3 can be explained as follows. 
Once a core is formed by a shock as a density peak along a filament (or on a sheet),
part of the gas is moving along the filament (or on the sheet) away from
the core, because of the pressure gradient. However, as long as the core becomes
massive enough, its gravitational force becomes larger than the pressure gradient
and the flow along the filament is reversed back into the core. In the simple
case of a single shock the collapsing core will be accreting along two 
filaments (the two sides of the original filament), while in the case of
the brightest cores, located at the intersection of a number of shocks,
many accreting filaments can be found.

\ifnum\inlinefig=1
\begin{figure}[!th]
\leavevmode
{\epsfxsize=5.45cm \epsfbox{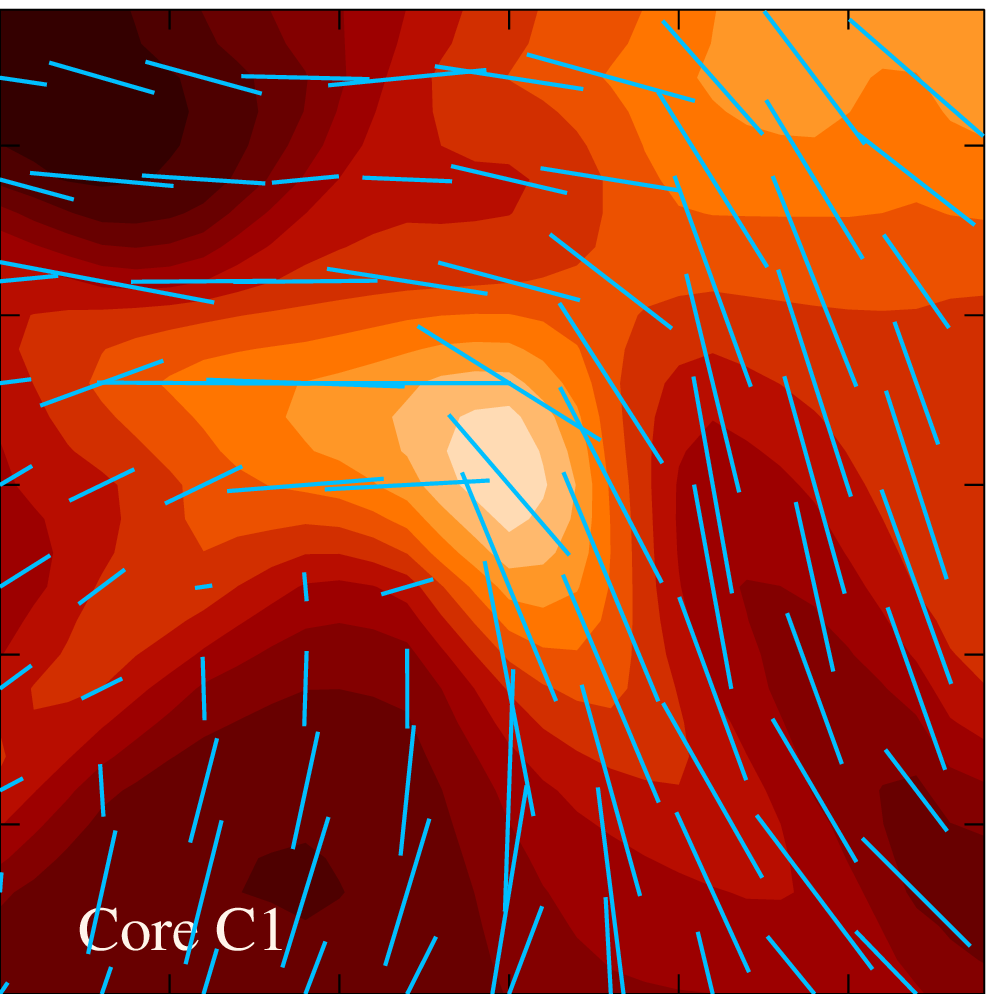}}
{\epsfxsize=5.45cm \epsfbox{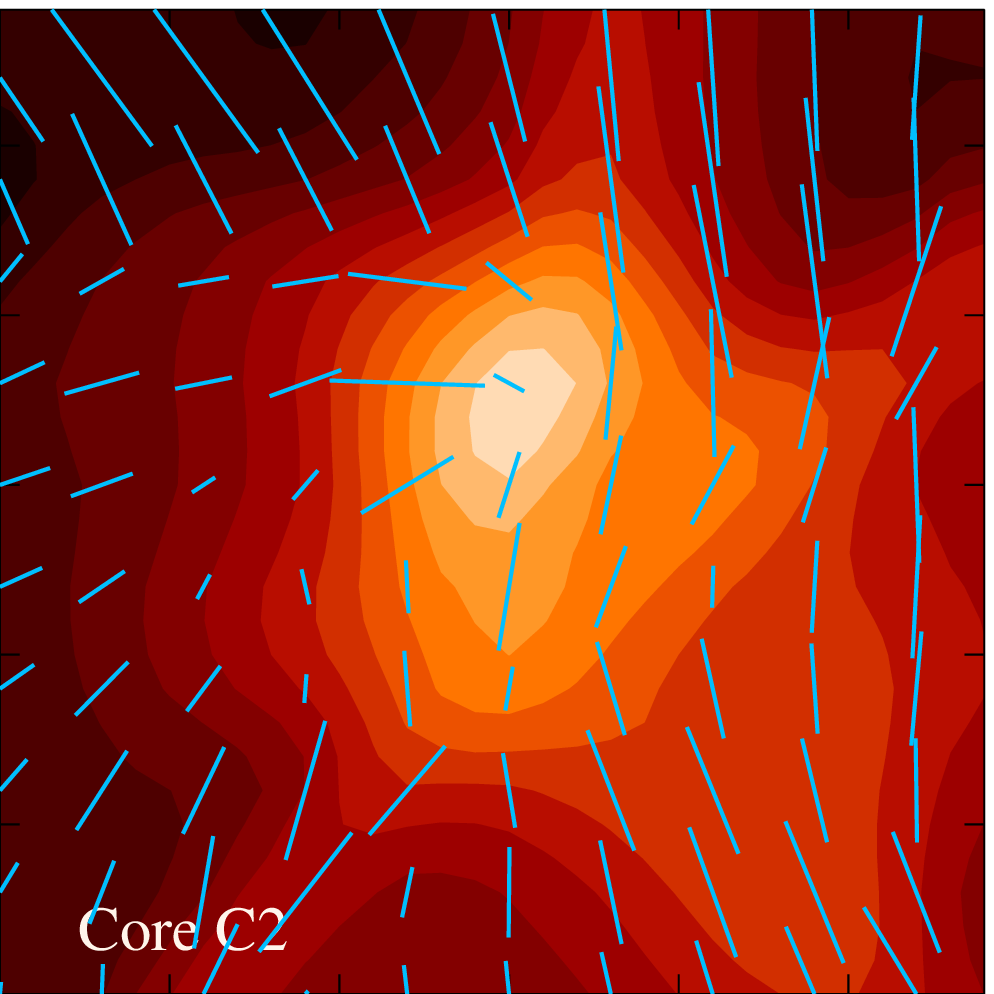}}
{\epsfxsize=5.45cm \epsfbox{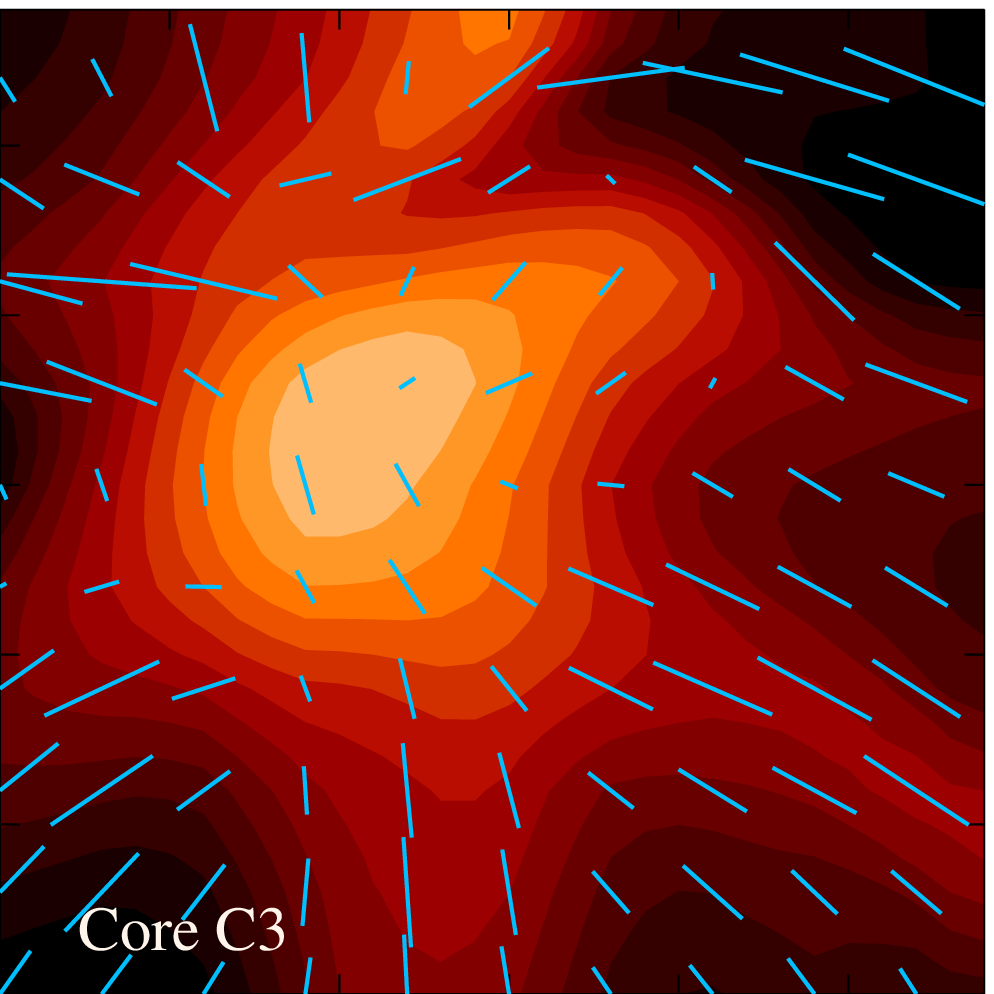}}
{\epsfxsize=5.45cm \epsfbox{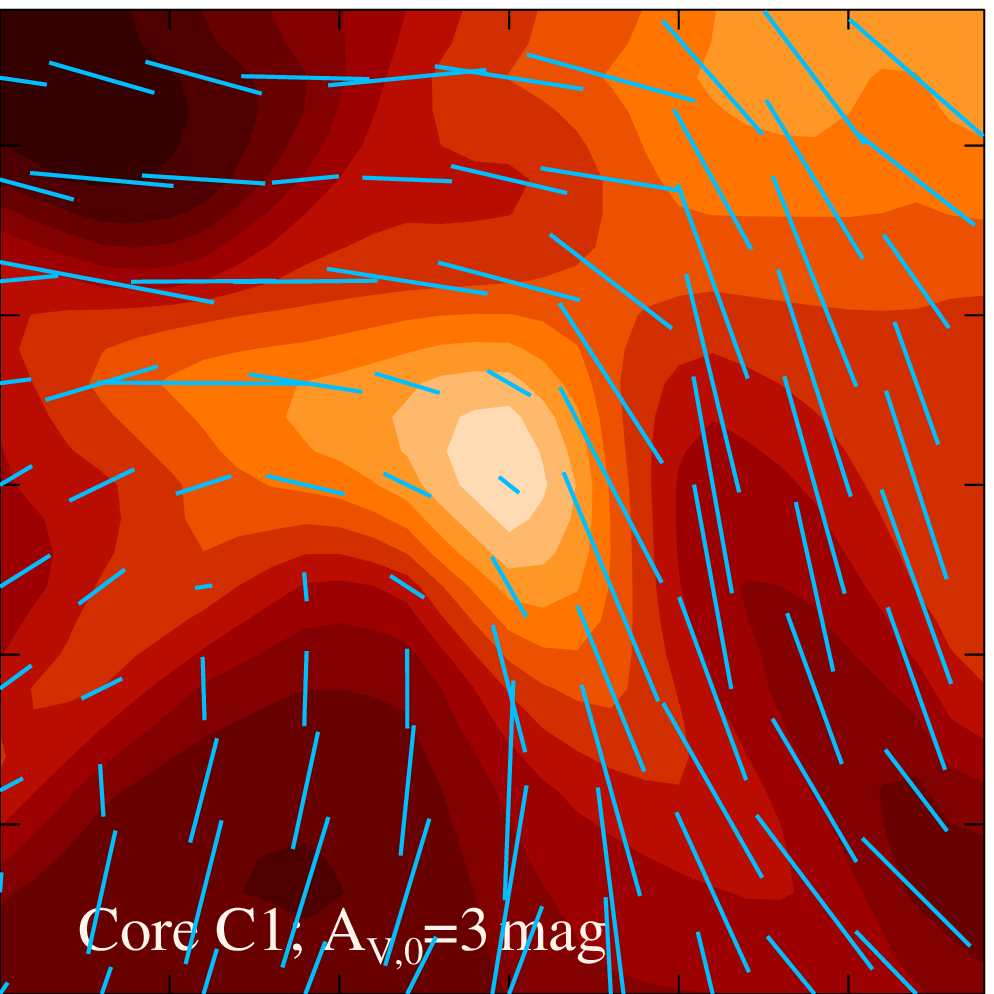}}
{\epsfxsize=5.45cm \epsfbox{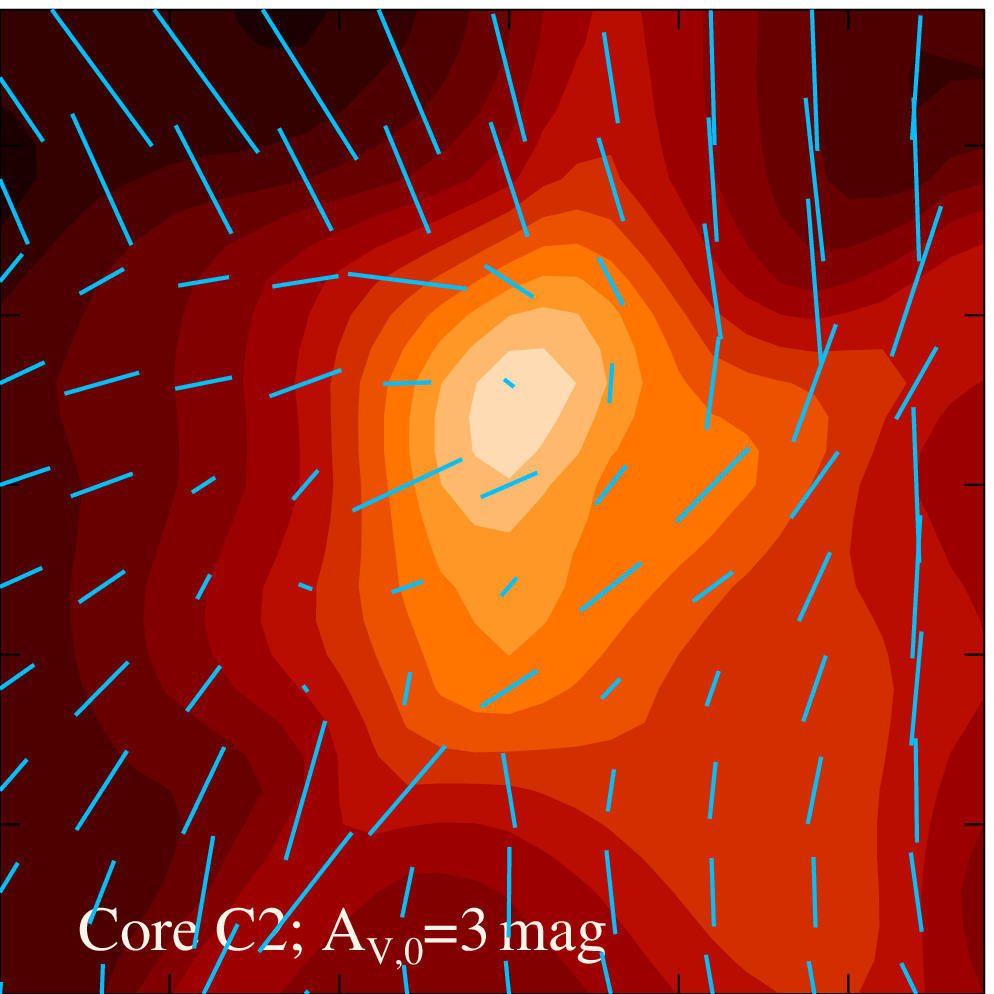}}
{\epsfxsize=5.45cm \epsfbox{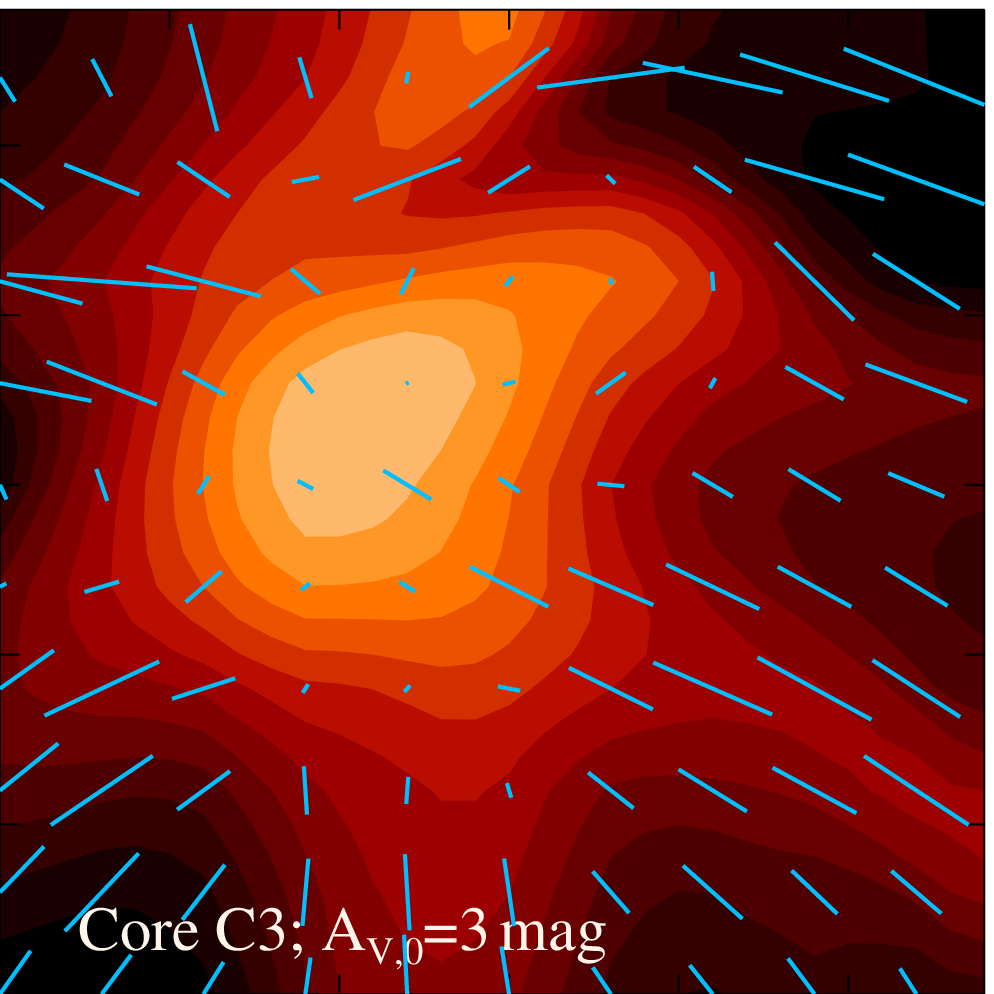}}
\caption[]{%
Upper panels: Polarization maps as in Figure~2, but limited to the regions around 
the cores C1, C2 and C3 (see text). The polarization vectors are 
here plotted for one position every two computational cells, and 
the largest polarization vectors (found in the core C1) corresponds 
to $P\approx 11.5$~\%. Lower panels: As upper panels, but from the map in Figure~3,
that is with unaligned grains above $A_{V,0}=3$~mag.}
\label{fig5}
\end{figure}
\fi

\ifnum\inlinefig=1
\begin{figure}[!th]
\leavevmode
{\epsfxsize=7.9cm \epsfbox{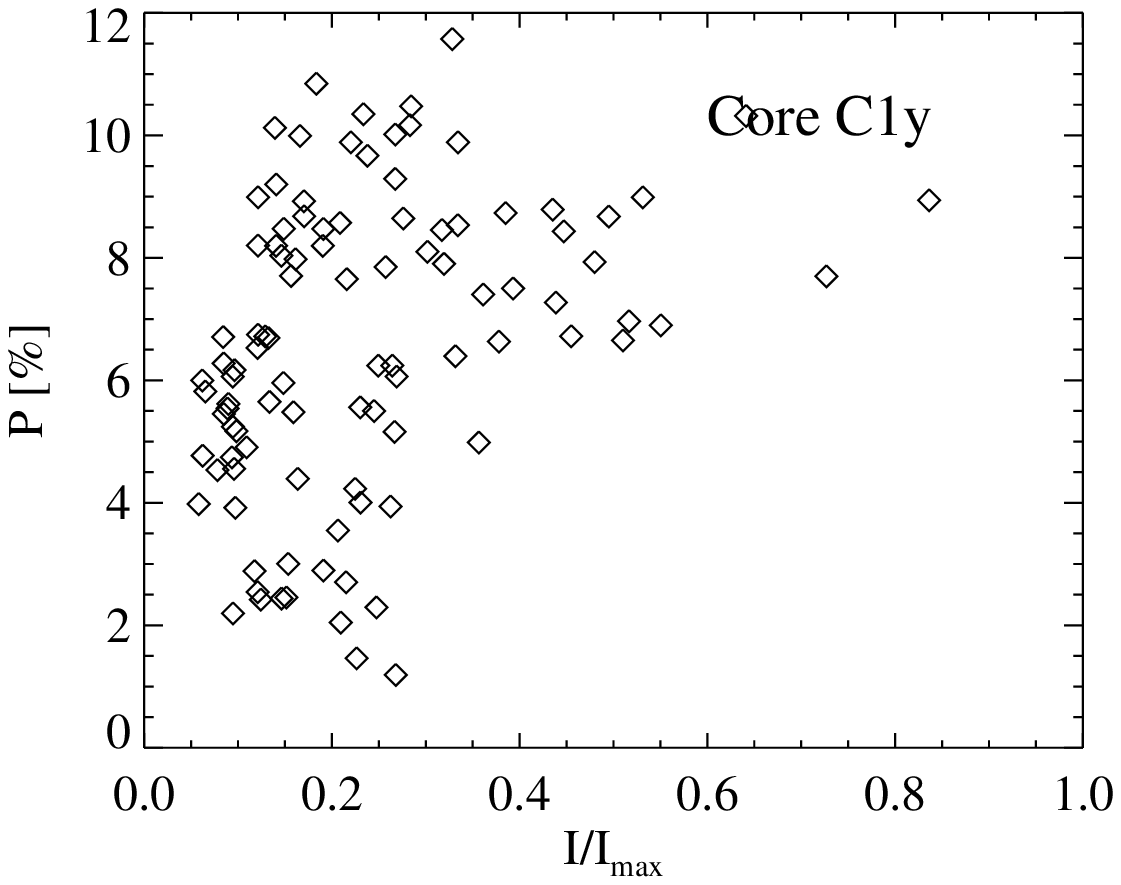}}
{\epsfxsize=7.9cm \epsfbox{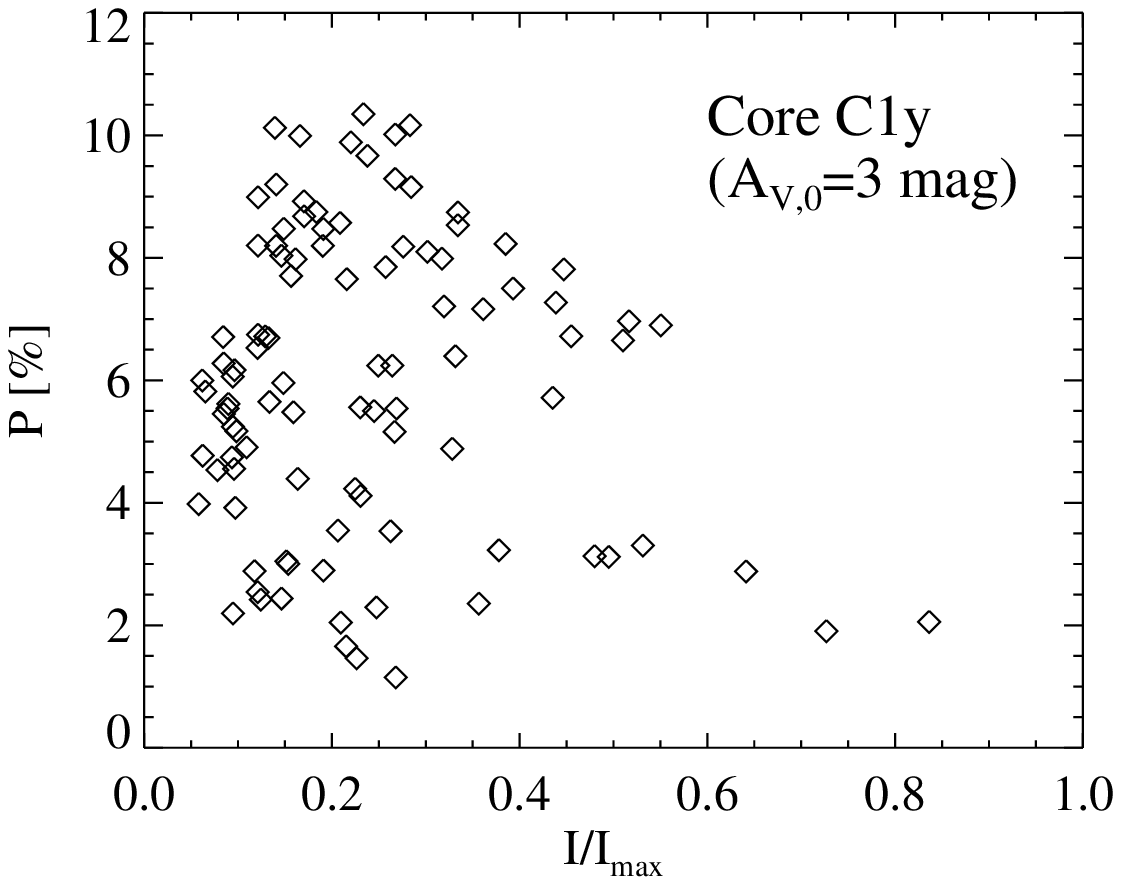}}
{\epsfxsize=7.9cm \epsfbox{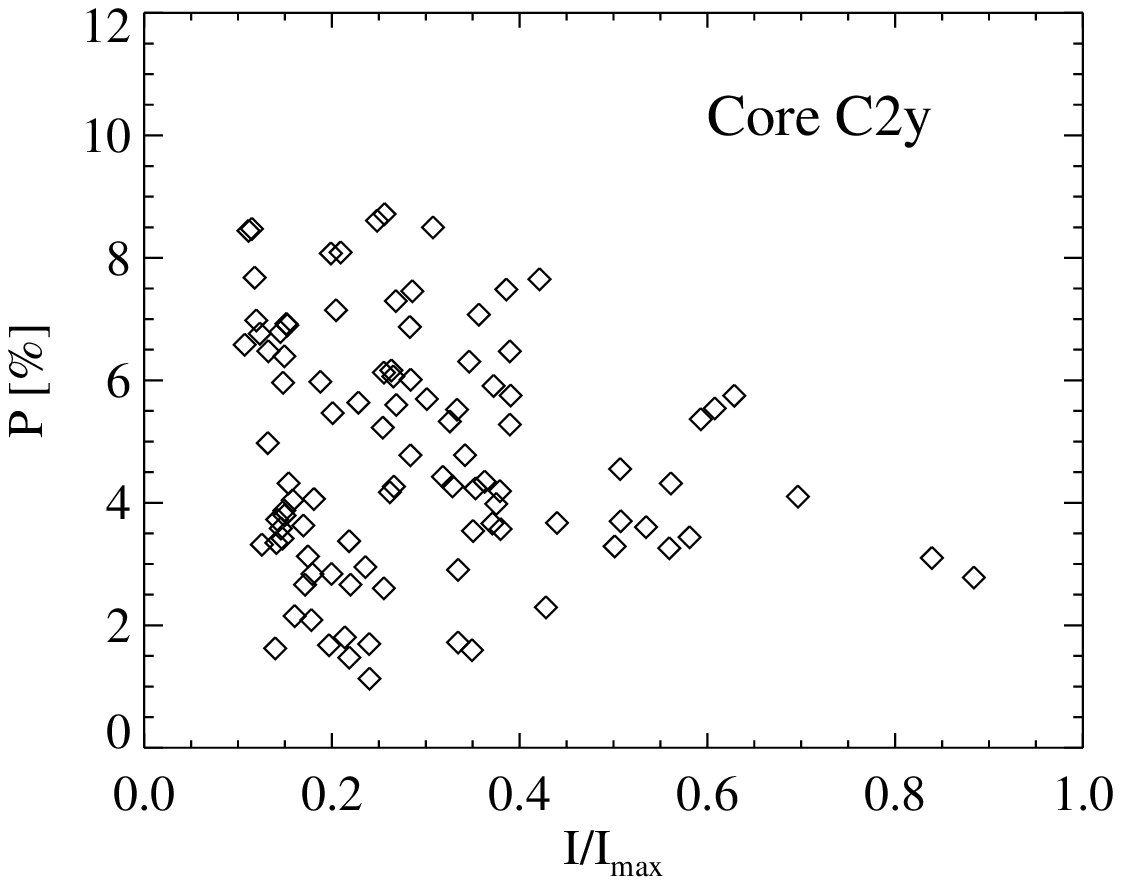}}
{\epsfxsize=7.9cm \epsfbox{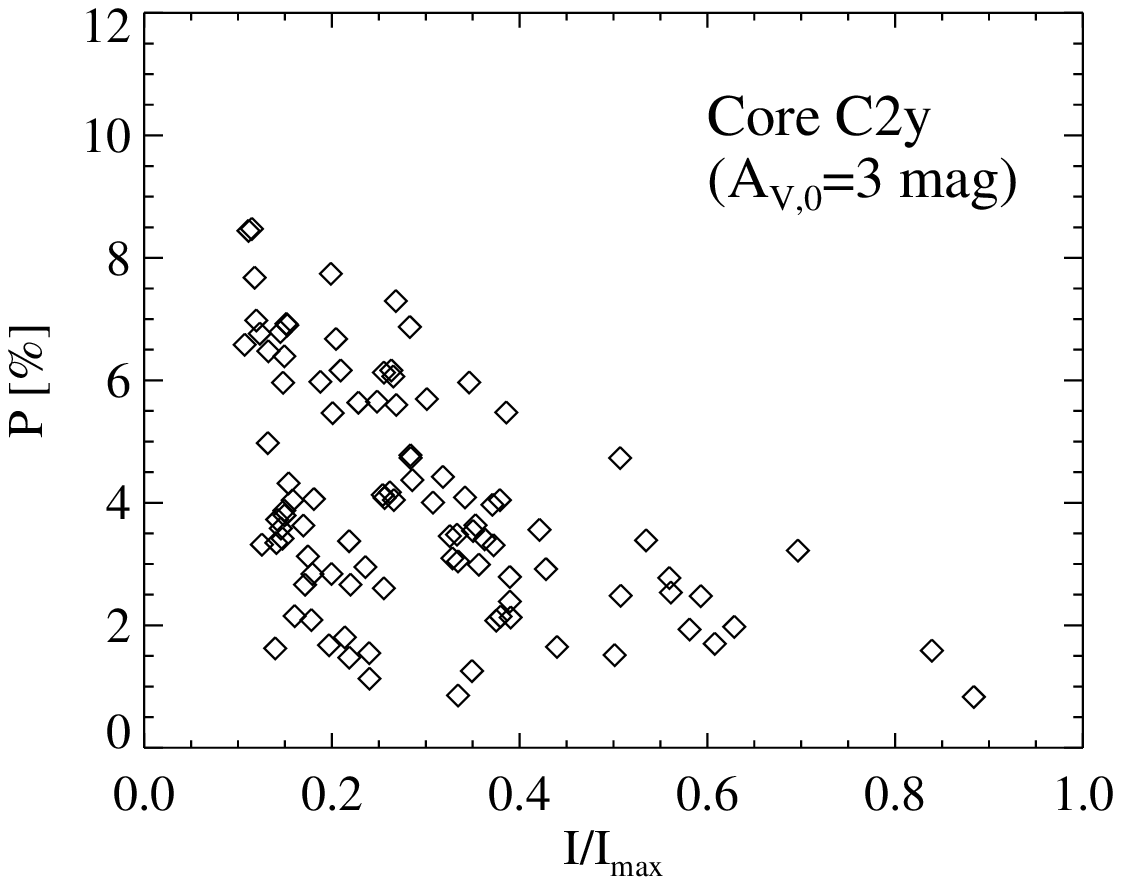}}
{\epsfxsize=7.9cm \epsfbox{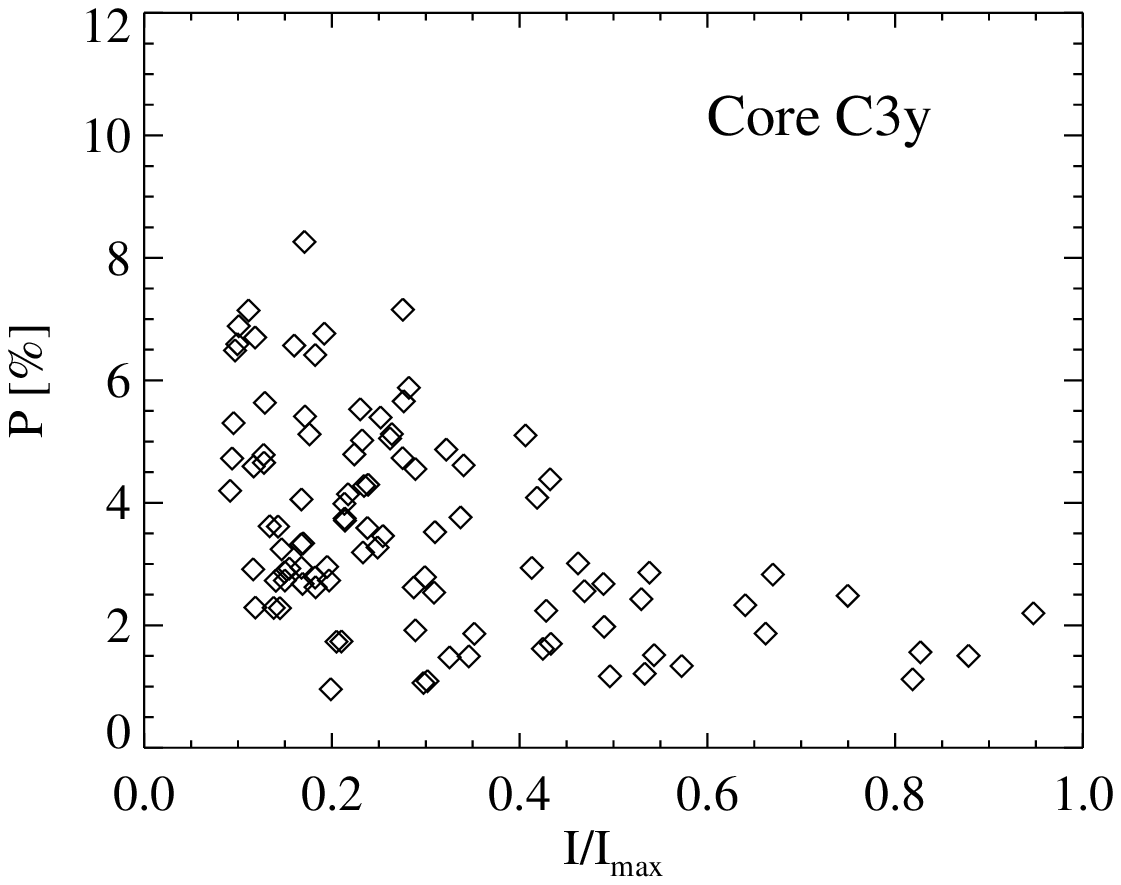}}
{\epsfxsize=7.9cm \epsfbox{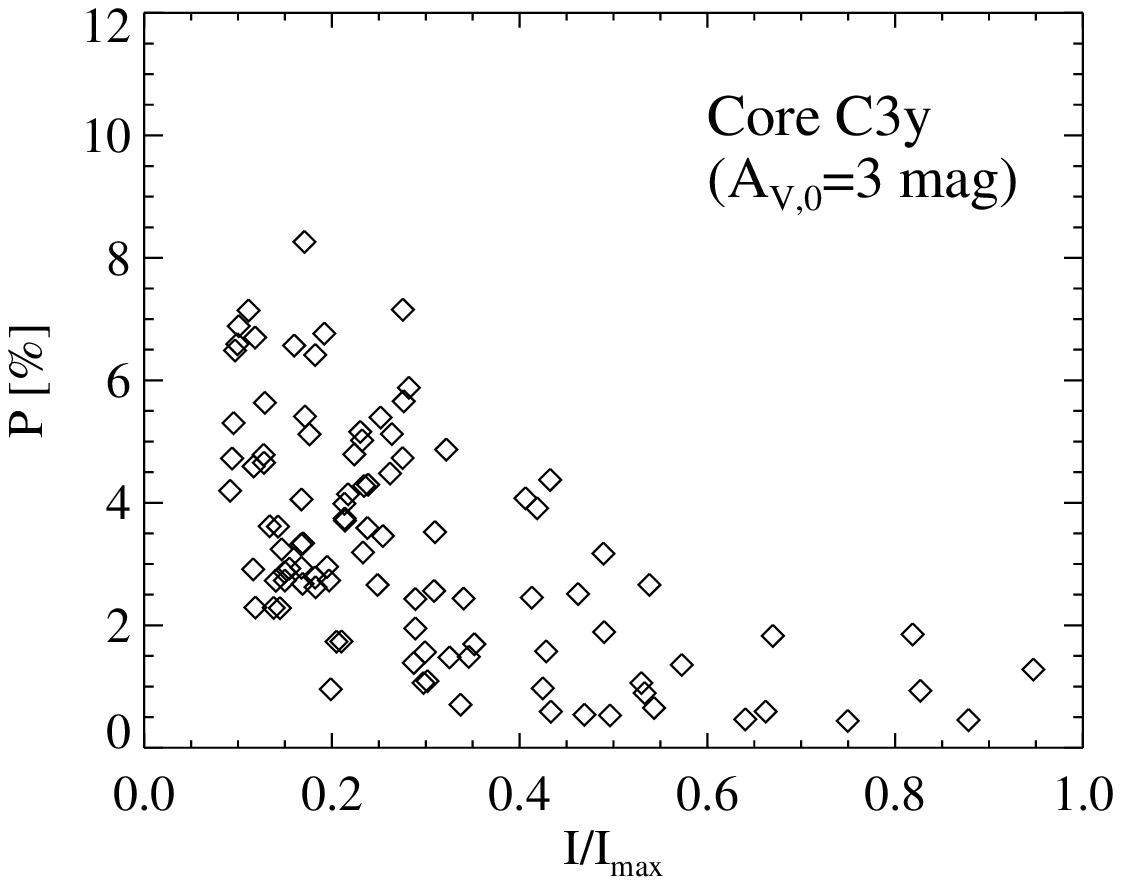}}
\caption[]{%
Left panels (model $A$): Scatter plots of the degree of polarization versus the emission intensity
from the maps in Figure~5 (upper panels). Only the polarization vectors shown in 
Figure~5 are used, that is one position every two computational cells. Right panels (model $B$):
same as left panels, but from the maps in the lower panels of Figure~5,
that is assuming unaligned grains above $A_{V,0}=3$~mag.}
\label{fig6}
\end{figure}
\fi

\ifnum\inlinefig=1
\begin{figure}[!th]
\leavevmode
{\epsfxsize=8.1cm \epsfbox{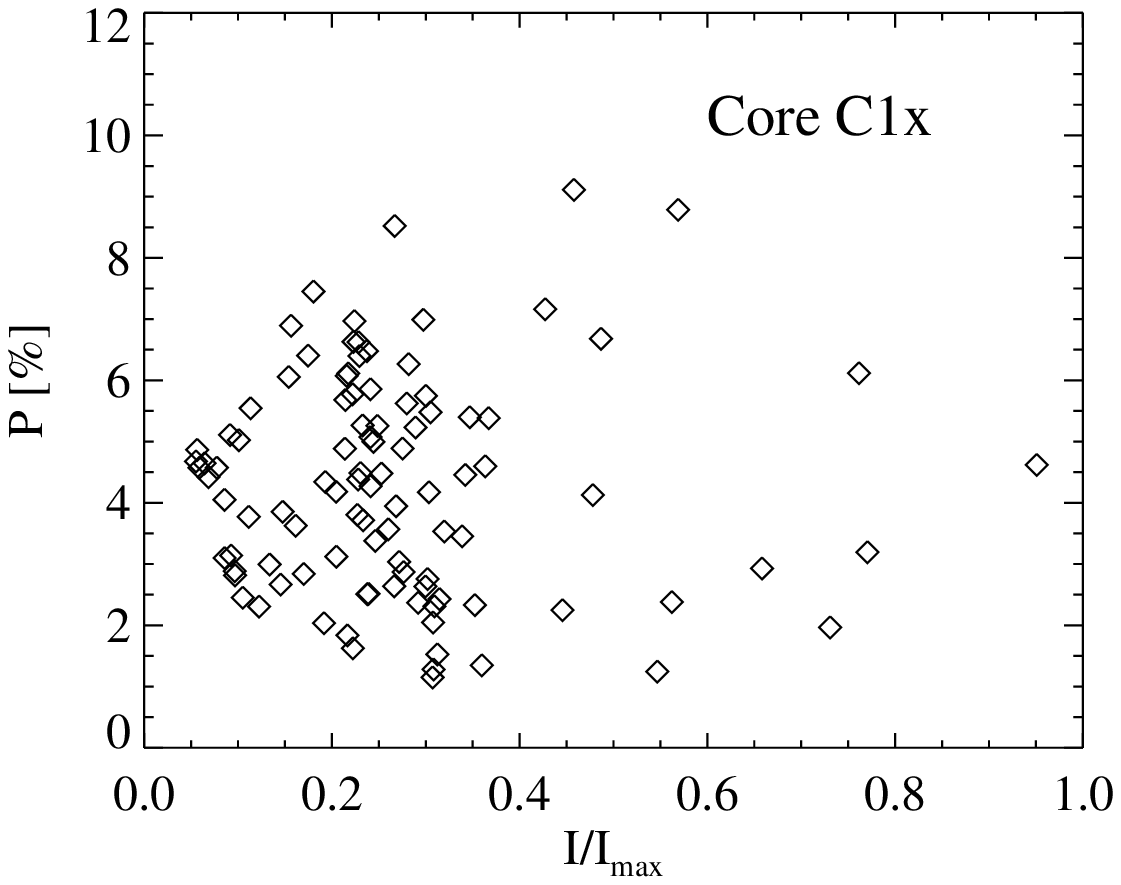}}
{\epsfxsize=8.1cm \epsfbox{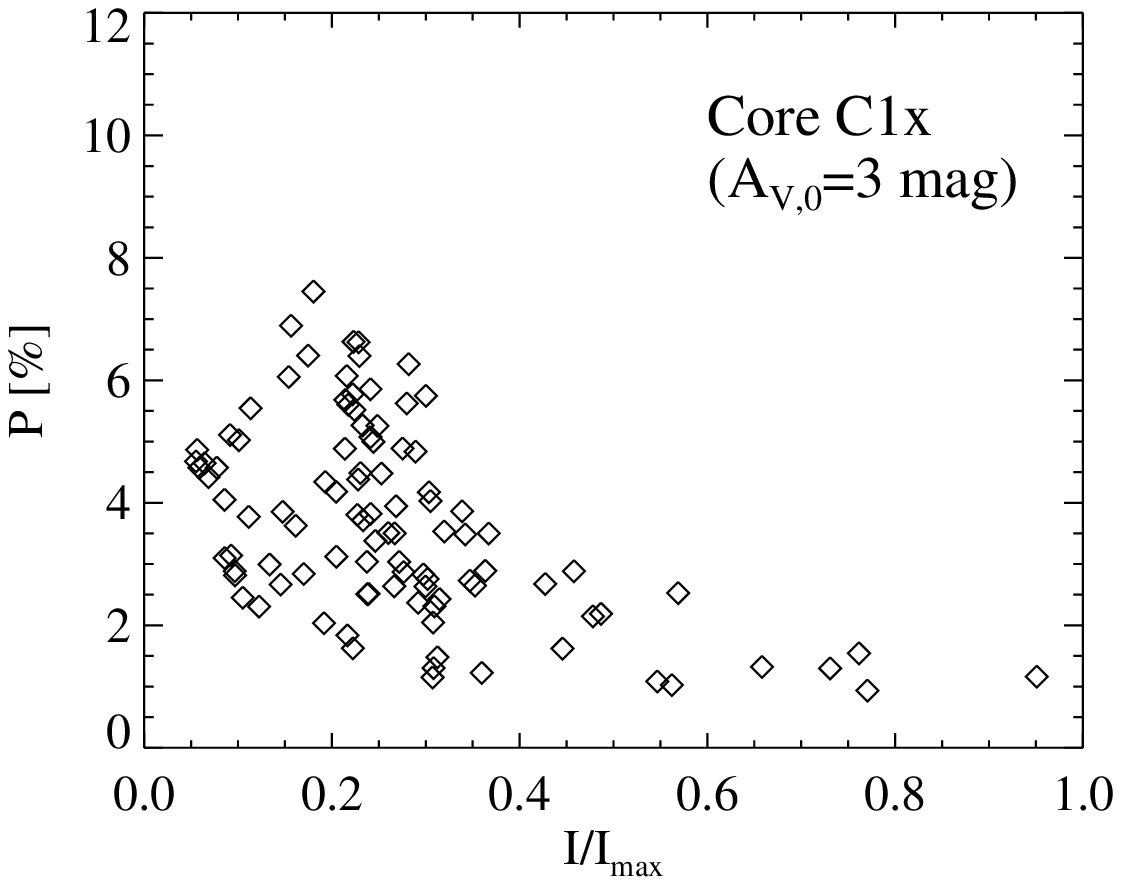}}
{\epsfxsize=8.1cm \epsfbox{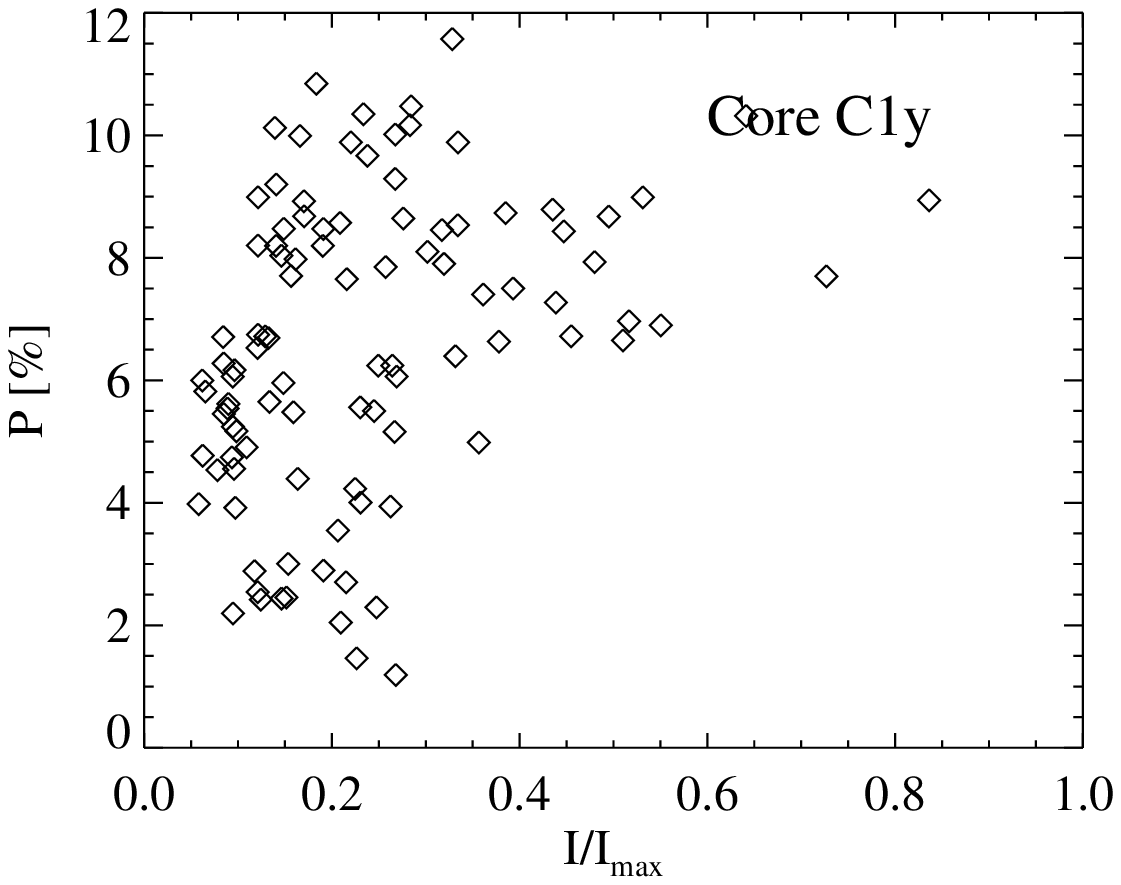}}
{\epsfxsize=8.1cm \epsfbox{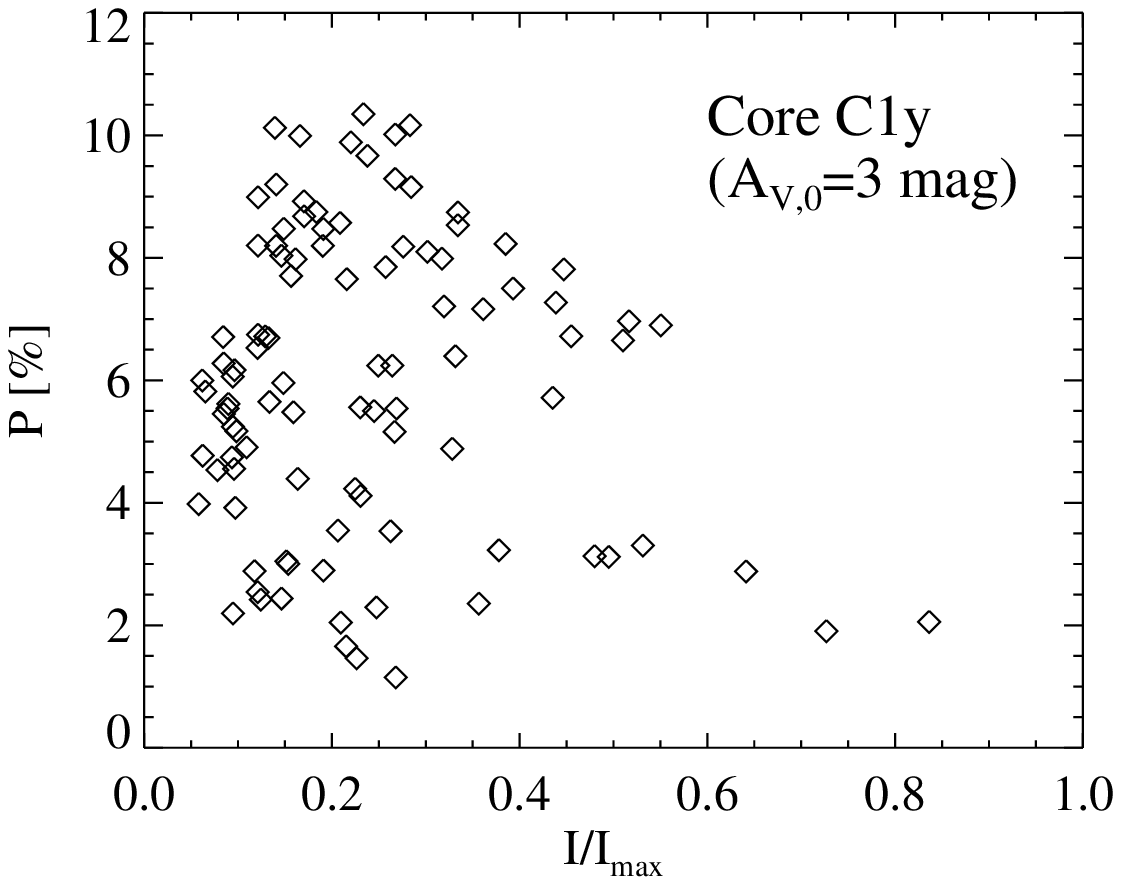}}
{\epsfxsize=8.1cm \epsfbox{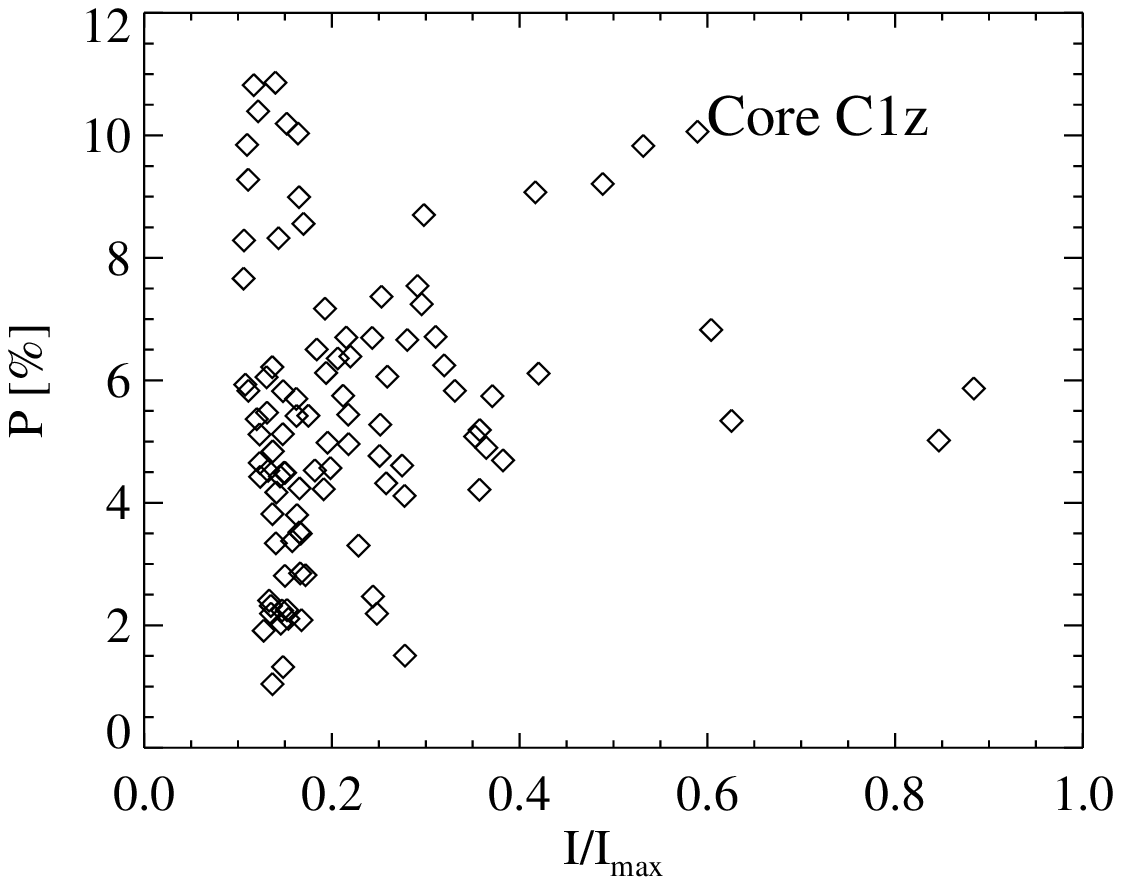}}
{\epsfxsize=8.1cm \epsfbox{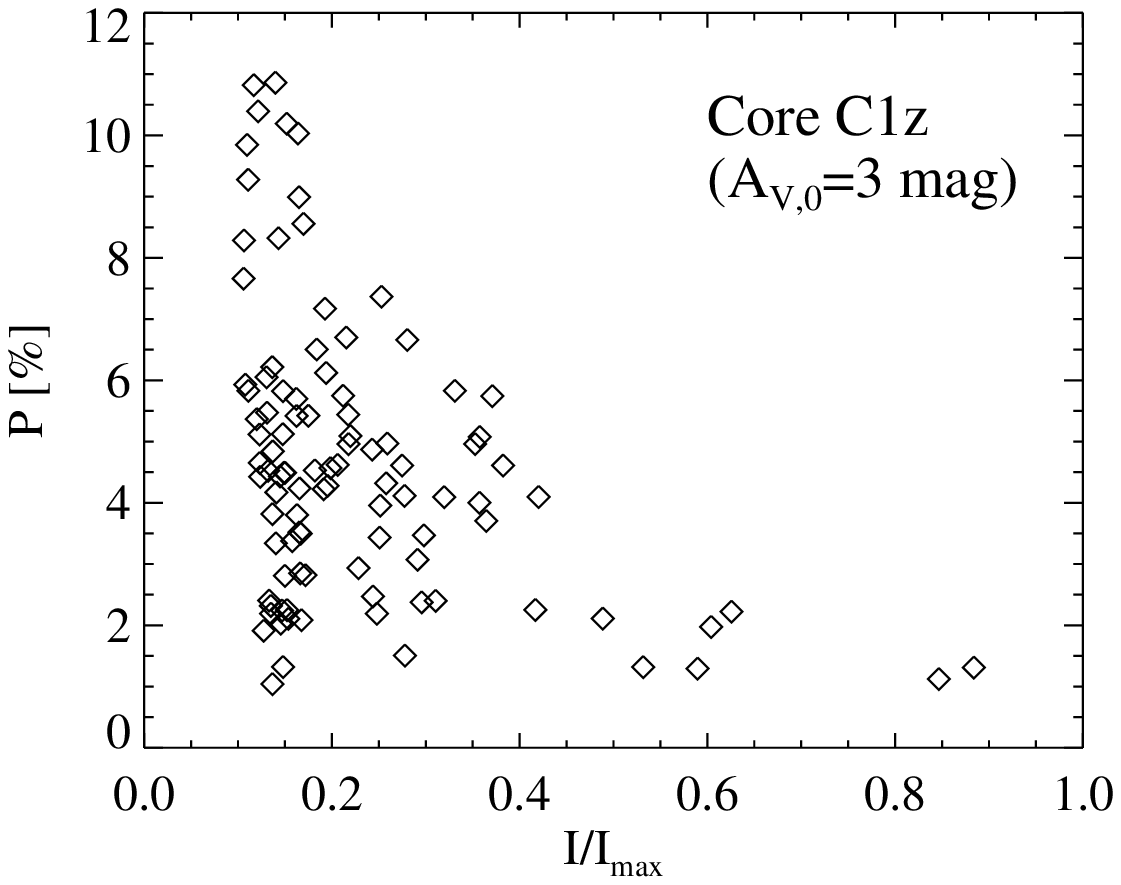}}
\caption[]{%
Left panel (model $A$): Scatter plot of the degree of polarization versus the emission 
intensity for the core C1, seen from three orthogonal points of view. Right panels (model $B$):
same as left panels, but with unaligned grains above $A_{V,0}=3$~mag.}
\label{fig7}
\end{figure}
\fi

The region of the polarization map around each core is shown in
the upper panels of Figure~5 for model $A$. The polarization vectors 
are here plotted for one position every two computational cells, 
and the largest polarization vectors (found in the core C1) corresponds 
to $P\approx 11.5$~\%. The respective scatter plots of $P$ versus $I$ 
(also for model $A$) are shown in the left panels of Figure~6, where the 
intensity $I$ is plotted in linear units, for a more direct comparison 
with published observational plots. The difference in the $P$--$I$ plot 
between the cores C1 and C3 is remarkable in model $A$. The plot for the 
core C1 is characterized by a lower envelope
with $P$ increasing with increasing $I$, while the plot for the core C3
shows a very clear upper envelope with $P$ decreasing with increasing
$I$. The excursion of the values of $P$ on the envelopes is from 
approximately 2\% to 9\% in the first case and from approximately
8\% to 2\% in the second case. The plot for the core C2 looks like
an intermediate case.

Should we conclude that the $P$--$I$ plot is a diagnostic 
for the time evolution of a core, from an early {\it shock phase}
to a {\it collapse phase}? The integration of the Stokes parameters
along different lines of sight shows that the $P$--$I$ plot of the
core C3 depends on the direction of the line of sight. Along certain
directions $P$ and $I$ are virtually uncorrelated; along others 
$P$ decreases with increasing $I$ because the magnetic field
is roughly parallel to the line of sight towards the center of the core,
or because of a cancellation effect due to the direction of polarization 
within material in front of the core. The $P$--$I$ plot is therefore
sensitive to the viewing angle, and is not a diagnostic 
for the evolutionary state of the cores (at least with the present numerical 
resolution that allows only the description of the very early phase of the
core collapse).

The situation is different when grains are assumed to be aligned only
at visual extinction below $A_{V,0}=3$~mag, as in model $B$. The polarization
maps of the cores in the case of model B are shown in the lower panels
of Figure~5, and the corresponding $P$--$I$ plots in the right panels
of Figure~6. The upper envelope in the $P$--$I$ plot, with $P$ decreasing with
increasing $I$, is always found,
since at visual extinction above $A_{V,0}=3$~mag the dust contributes
to the continuum intensity $I$, but not to the polarized flux ($P$ is
defined as the ratio of polarized and total flux --eq. (\ref{pdef})).
The direction of polarization is also affected by the lack of alignement
of grains at large visual extinction. As an example, the region at the south
of the core C2 shows a magnetic field direction roughly parallel
to the elongation of the core in model $A$, while the magnetic field
direction is almost perpendicular to the core elongation in model $B$.

Figure~7 shows the $P$--$I$ plots for the core C1 seen from three
different orthogonal directions, in model $A$ (left panels) and
$B$ (right panels). These plots show that $P$ always decreases with 
increasing $I$ in model $B$, independent of the viewing angle.

The histograms of the polarization angle $\chi$ in the maps of
Figure~5 are plotted in Figure~8. For computing these histograms we have
used all the polarization vectors available, that is one per computational 
cell. Since the maps are made of $20\times 20$ pixels, we have used 400
polarization vectors for each histogram, while in Figures~4 and 5 only
$10\times 10=100$ vectors were used for clarity. The plots in Figure~7 show very broad 
distributions of $\chi$. The values of $\chi$ are limited to the
interval $[-\pi/2,\pi/2]$, but the reference direction, or the origin 
of the interval, can be shifted arbitrarily by an amount between 0 and $\pi$.
We have done so in order to minimize the dispersion of $\chi$, which is
obtained by shifting the origin of the angle to the approximate location
of the minimum in the histogram. In this way the dispersion $\sigma_{\chi}$
has a robust physical meaning, indipendent of the particular local orientation
of the average magnetic field relative to the North. No Gaussian fit of the
histogram is attempted, and $\sigma_{\chi}$ is simply defined as
\begin{equation}
\sigma_{\chi} = \langle (\chi - \langle \chi \rangle)^2\rangle^{1/2}  
\label{sigmachi}
\end{equation}
with the values of $\chi$ defined in the interval that minimizes this
expression, as explained above. Differences between models $A$ and $B$ are
rather small. Results are listed Table~1. For both models $A$ and $B$ we 
find the average value of the rms $\chi$ to be $\sigma_{\chi}=0.60\pm 0.11$.

\ifnum\inlinefig=1
\begin{figure}[!th]
\leavevmode
{\epsfxsize=7.9cm \epsfbox{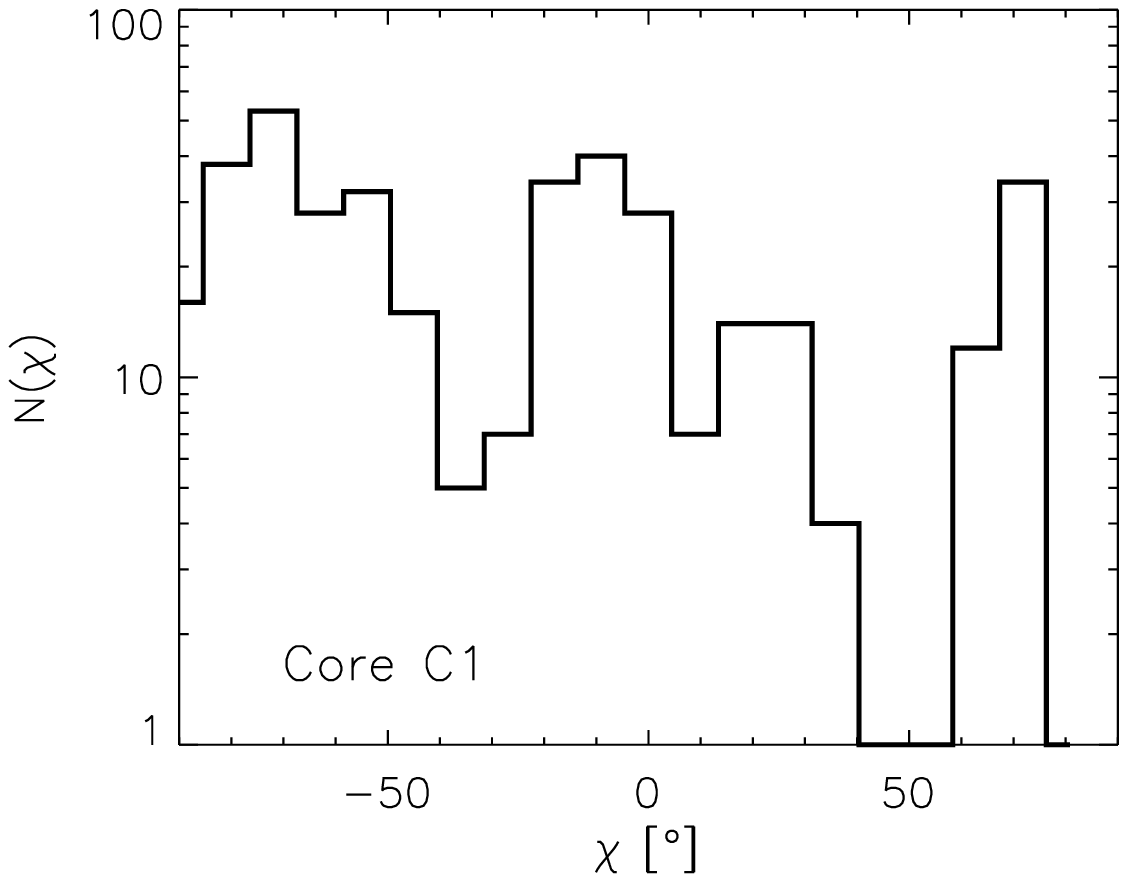}}
{\epsfxsize=7.9cm \epsfbox{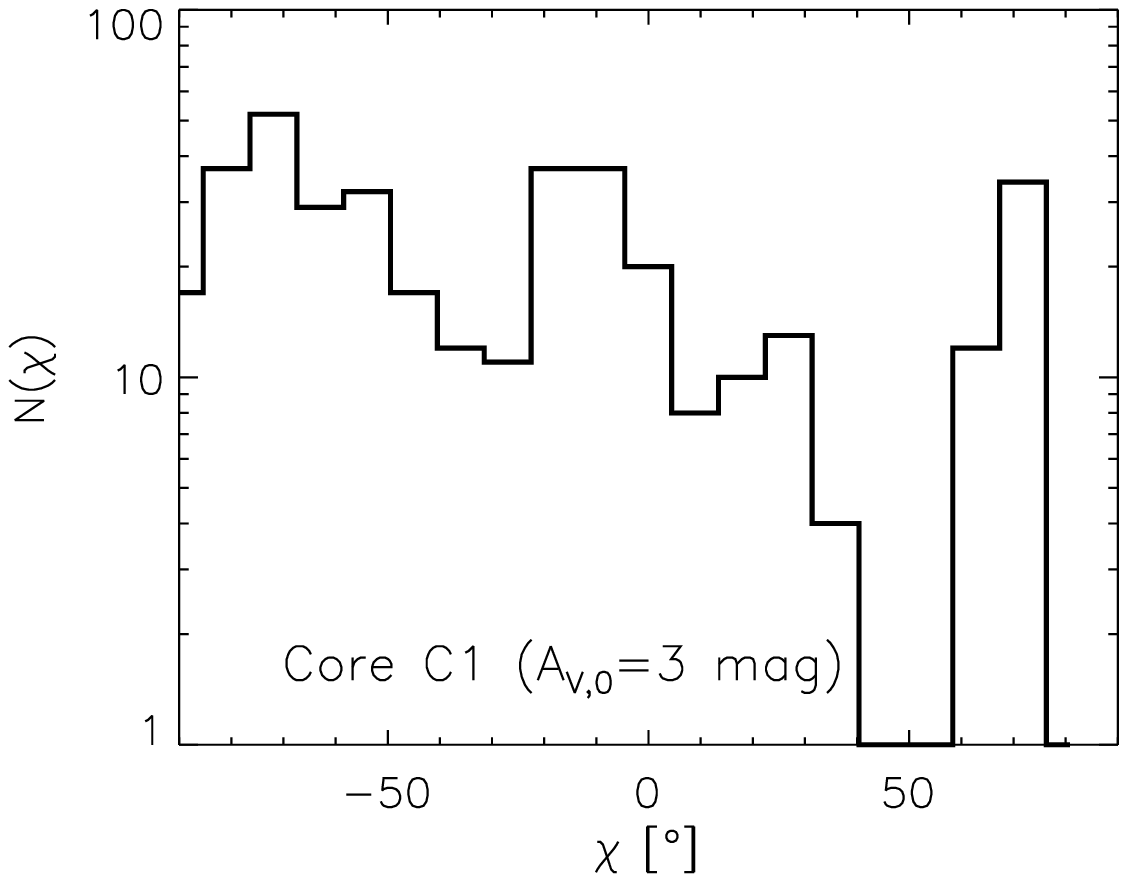}}
{\epsfxsize=7.9cm \epsfbox{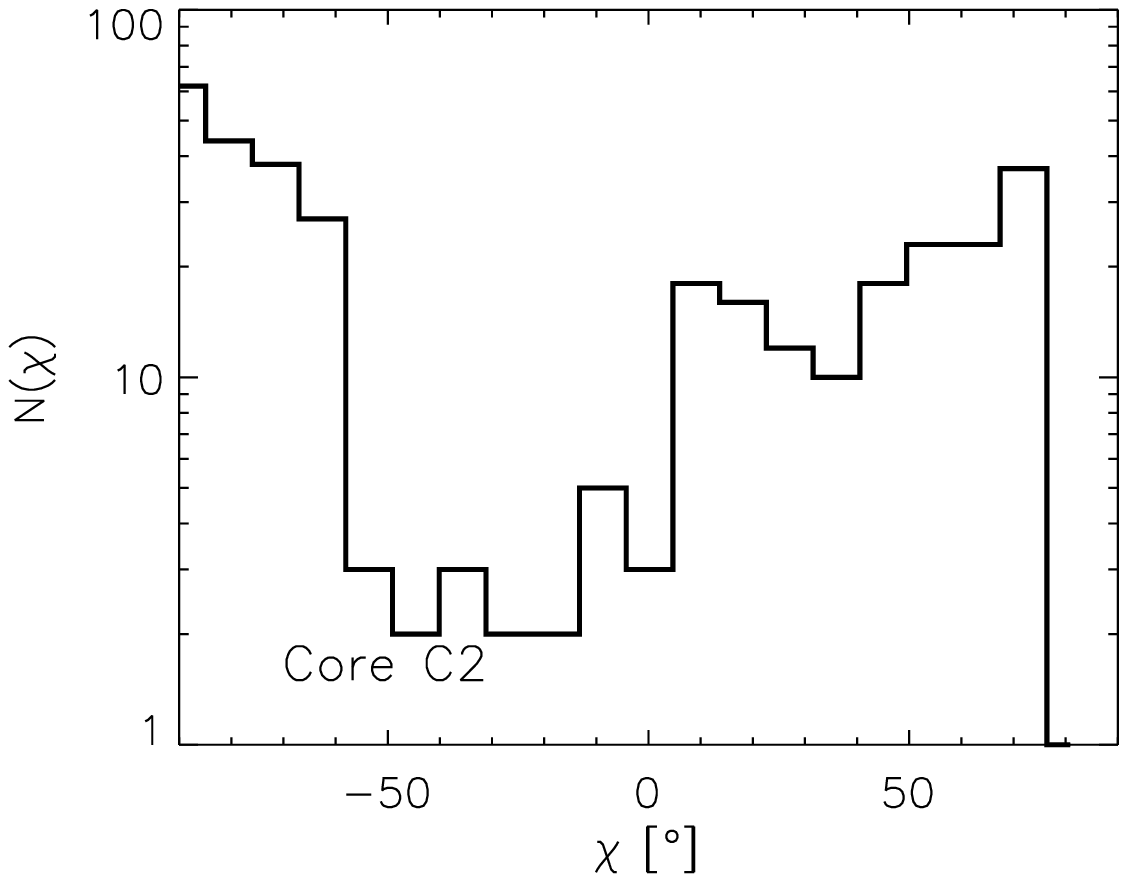}}
{\epsfxsize=7.9cm \epsfbox{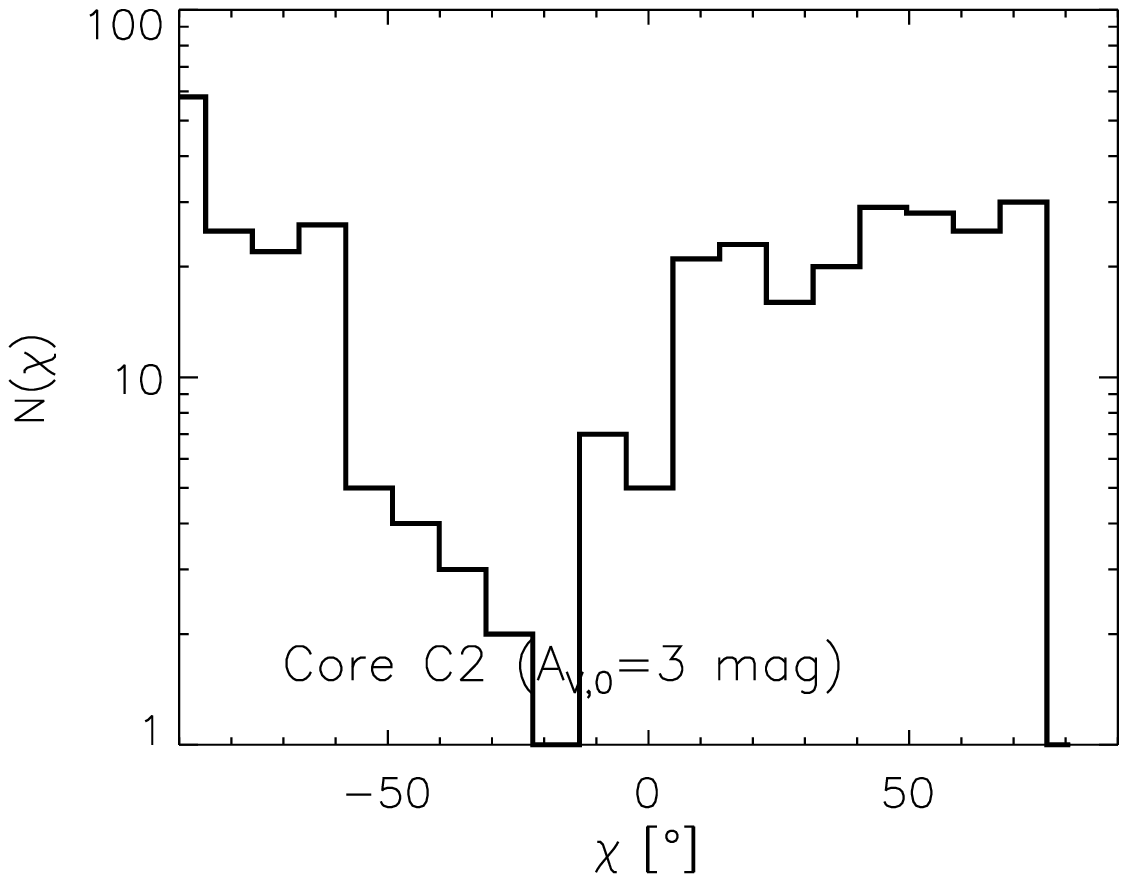}}
{\epsfxsize=7.9cm \epsfbox{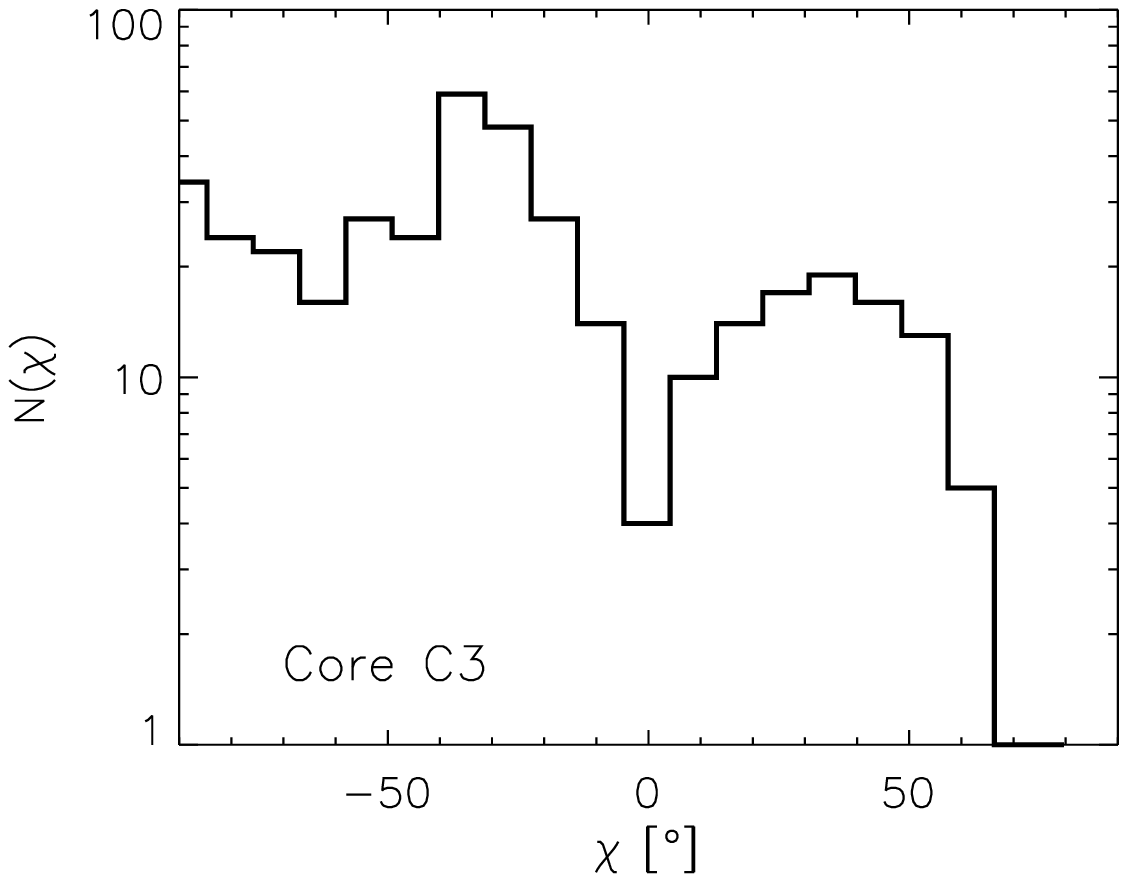}}
{\epsfxsize=7.9cm \epsfbox{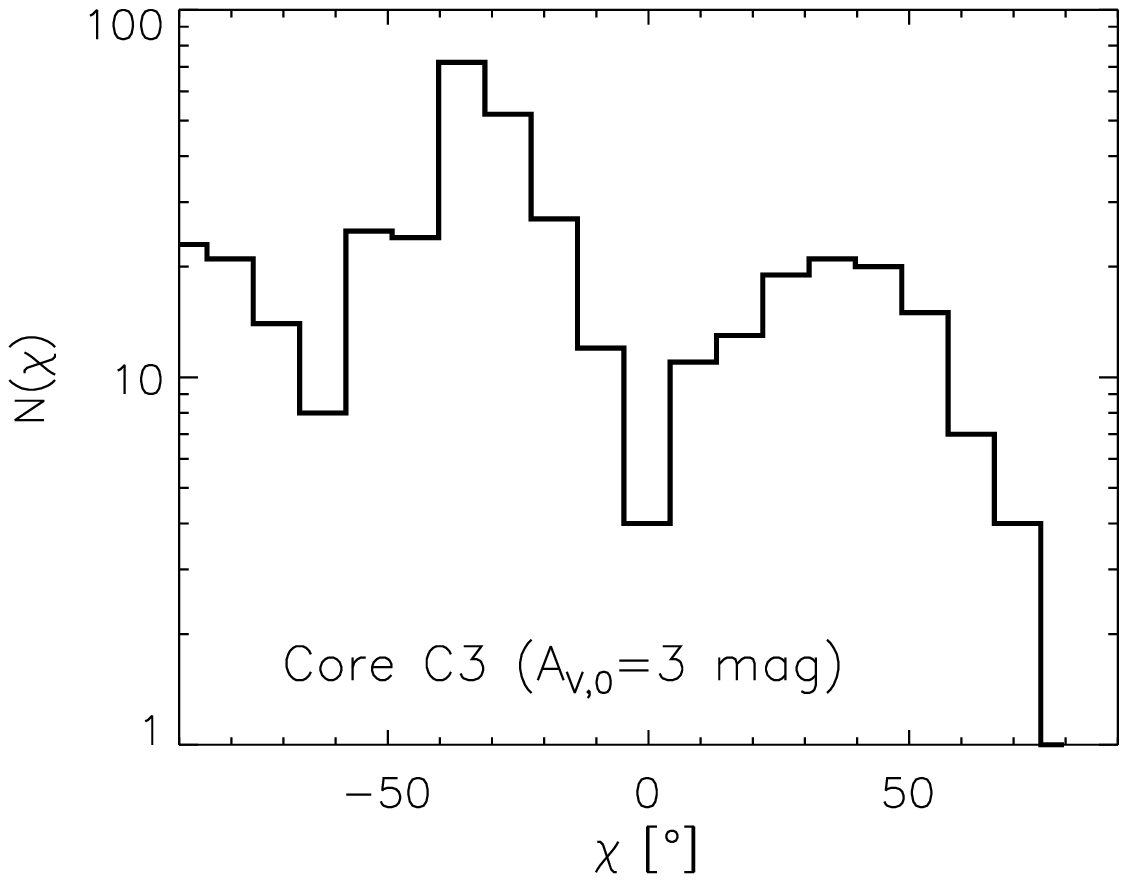}}
\caption[]{%
Left panels (model $A$): Histograms of the polarization angle $\chi$ from the polarization maps 
shown in Figure~5 (upper panels). All the available polarization vectors have been used, 
that is one per computational cell. Right panels (model $B$): same 
as left panels, but from the polarization maps in the lower panels of Figure~5, that is
with unaligned grains above $A_{V,0}=3$~mag.}
\label{fig8}
\end{figure}
\fi

\ifnum\inlinefig=1
\begin{figure}[!th]
\leavevmode
{\epsfxsize=5.4cm \epsfbox{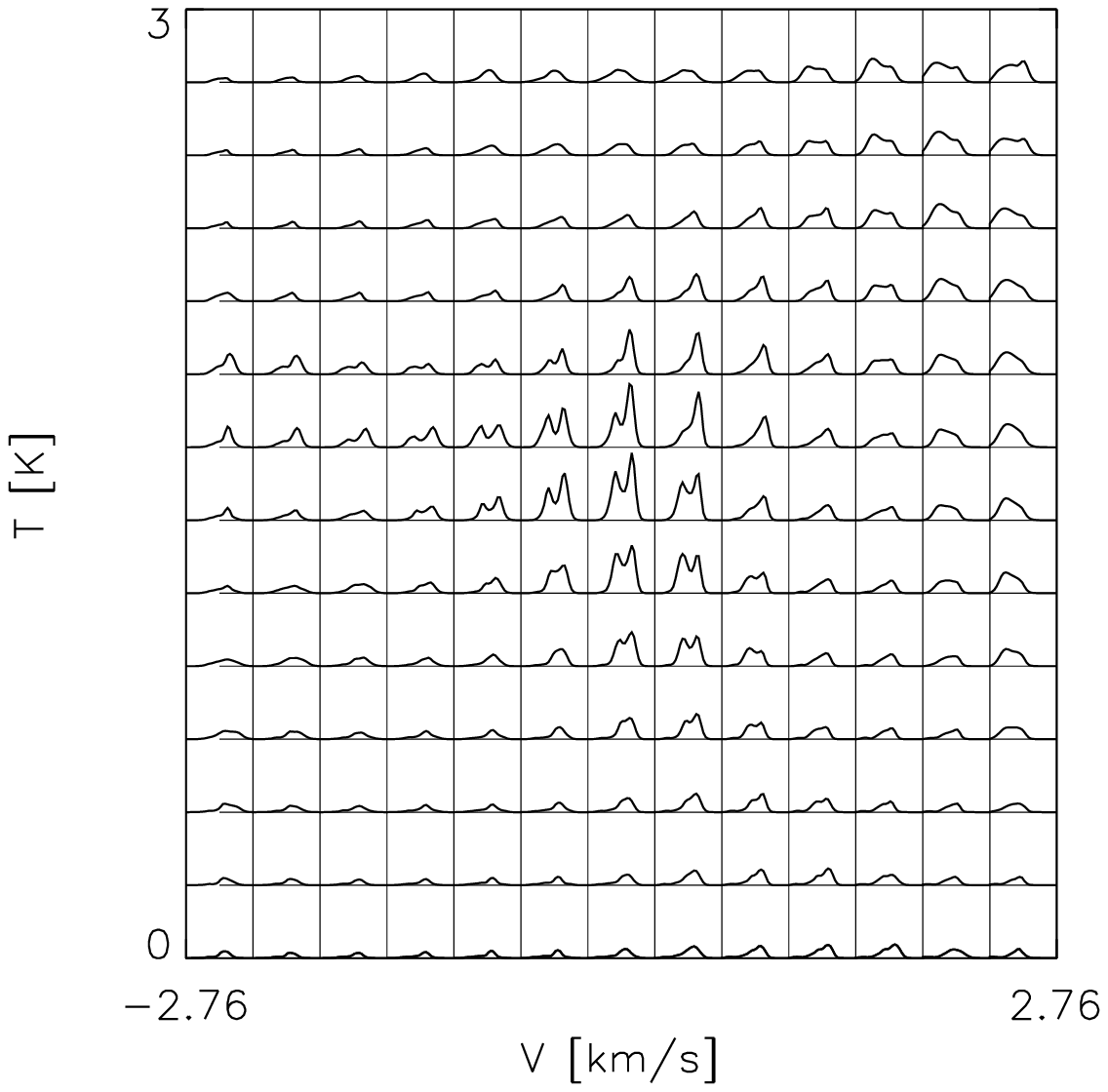}}
{\epsfxsize=5.4cm \epsfbox{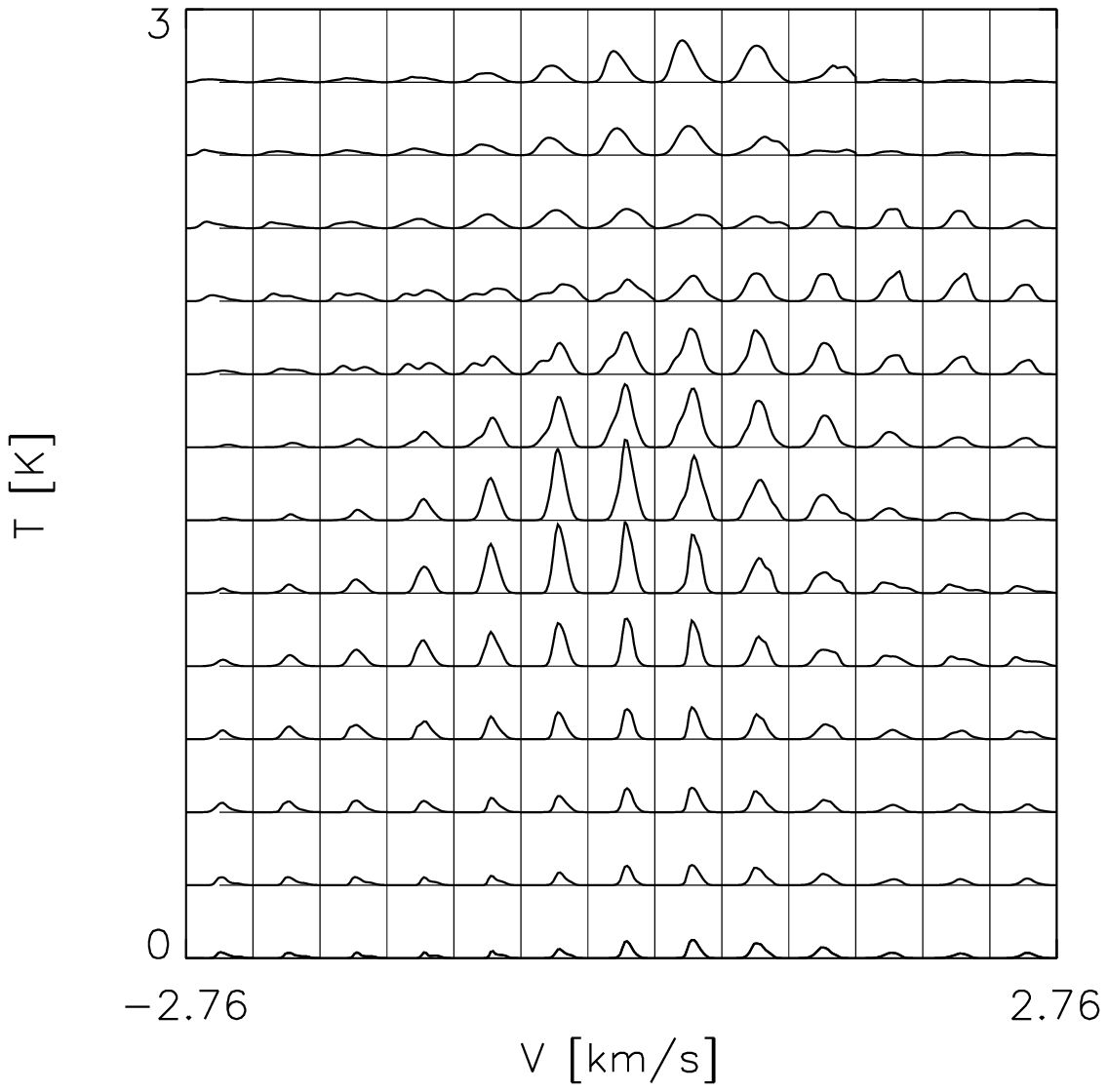}}
{\epsfxsize=5.4cm \epsfbox{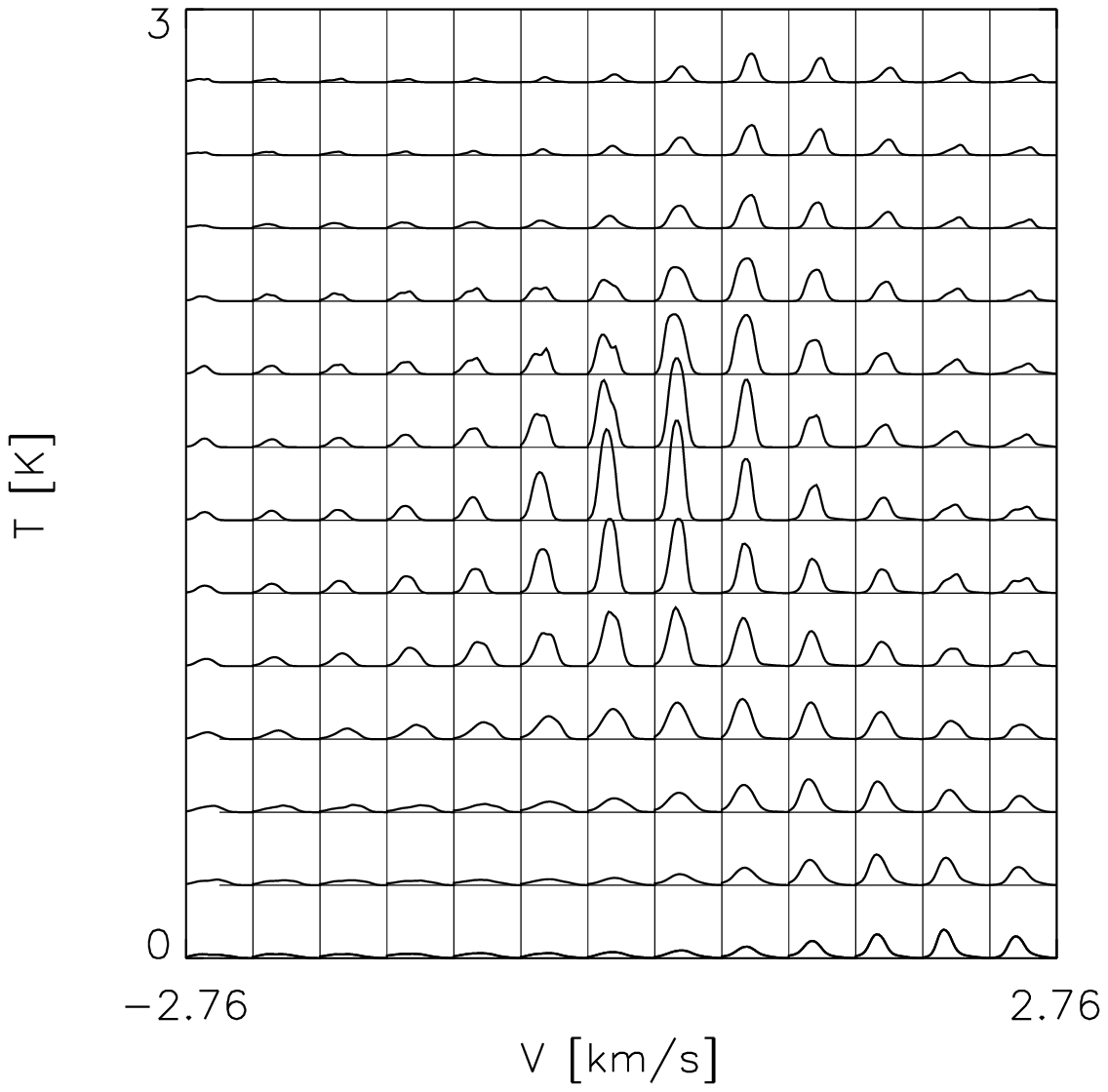}}
\caption[]{%
Spectral maps of the J=1--0 transition of CS. The maps are located
around the cores C1, C2 and C3, from top to bottom respectively. The regions
shown in the maps correspond to the regions included in the maps
of Figure~5. The velocity axis ranges from -2.76 to 2.76 km/s, and the antenna
temperature axis from 0 to 3 K.}
\label{fig9}
\end{figure}
\fi

\section{Observational $P$--$I$ Plots of ``Quiescent'' Cores}

In this work we define as ``quiescent'' any starless or low mass 
star forming core. We do not try to model regions of massive star 
formation, where the enhanced radiation field must have a considerable
effect on the grain alignment (Draine \& Weingartner 1996, 1997).

\nocite{Draine+Weingartner96}  \nocite{Draine+Weingartner97}

A number of recent sub-mm polarization maps of quiescent cores
can be used to study the correlation of the degree of polarization
$P$ and the sub-mm dust continuum intensity $I$. Davis et al. (2000)
observed the Serpens low mass star forming region, which contains
eight cores. Polarization maps of the three cores L1544, L183 and L43 were 
obtained by Ward--Thompson et al. (2000) and three more cores, CB 26, 
CB 24 and DC 253-1.6 (CG 30) were mapped by Henning et al. (2001). 
All these regions were observed at 850~$\mu$m, with the SCUBA bolometer
array at the James Clerk Maxwell Telescope. In all of the 14 cores
the value of $P$ is found to decrease rather steeply with increasing
$I$. Polarization vectors from the lower intensity regions around the cores
are always much larger than the vectors close to the center of the cores.

$P$--$I$ plots are presented both in Davis et al. (2000) and in 
Henning et al. (2001). They are very similar to the $P$--$I$
plots from our model $B$, presented in the previous section.
The upper envelope of the plots is defined by values of $P$ 
from 1-2\% at the highest $I$, to 8-10\% 
at the lowest values of $I$. The drop in $P$ is roughly 
consistent with a constant polarized flux, where the polarized 
flux is defined as $P\times I$. This suggests that a significant
fraction of the dust emission is from grains that do not contribute
significantly to the polarization, presumably unaligned grains.

Model $A$ cannot reproduce the same result for all cores, but only in
some cases due to the fortuitous orientation of the magnetic field
or to a cancellation effect also dependent on the orientation of the
core relative to the line of sight. It is possible that a numerical
simulation with higher resolution, able to follow the collapse of cores
to larger densities, may generate a stronger cancellation effect. The
gravitational collapse is expected to drag the lines of force of the magnetic 
field towards the core center, which could generate a cancellation effect.
However, it is not clear how important this effect may be, once diluted
over the size of the telescope beam. It is also difficult to imagine
it should be independent of orientation, since the average magnetic field
before the collapse defines a specific direction.

\section{Distribution of Polarization Angle and Magnetic Field Strength}

In linear transverse oscillations of a line of force (for example  
Alfv\'{e}n waves in a uniform medium) the ratio of the lateral
and transverse velocities is equal to the deviation of the line of
force from a straight line. This deviation of the field from a straight 
line is measured by the angle
$\chi$ between the straight line and the local line of force.
If the transverse velocity is the Alfv\'{e}n velocity $v_a$,
and the lateral velocity is the rms turbulent velocity $\sigma_v$,
then
\begin{equation}
\sigma_{\chi} \approx \frac{\sigma_v}{v_a} \equiv {\cal M}_a,
\label{chi1}
\end{equation}
where ${\cal M}_a$ is the Alfv\'{e}nic Mach number.
This is strictly valid only in linear theory. We can define a coefficient
$f$ that quantifies the deviation from the prediction of the linear theory:
\begin{equation}
f \equiv \frac{\langle v_a^2 \rangle^{1/2}}{\sigma_v/\sigma_{\chi}} 
    =  \frac{\langle B^2/(4\pi\rho) \rangle^{1/2}}{\sigma_v/\sigma_{\chi}},
\label{f}
\end{equation}
where the numerator is the rms Alfv\'{e}n velocity, averaged inside the
volume under study. Equation (\ref{chi1}), or equivalently the value $f=1$,
was first used to derive the magnetic field strength in the Galaxy by 
Davis (1951) and Chandrasekhar and Fermi (1953). 
This method was further discussed by Zweibel (1990) and Myers \& Goodman (1991). 
It has also been tested on the results of MHD simulations of sub--Alfv\'{e}nic 
turbulence by Ostriker, Stone \& Gammie (2001). They find that for small dispersion 
of the polarization angle ($\sigma_{\chi} < 0.44$ radians) the Chandrasekhar--Fermi 
formula provides a good estimate of the plane of the sky field strength, 
if a coefficient $f \approx 0.5$ is used. Similar results are found by Heitsch
et al. (2001), using numerical simulations of mildly super--Alfv\'{e}nic
turbulence.

\nocite{Davis51}  \nocite{Zweibel90} \nocite{Myers+Goodman91}
\nocite{Heitsch+2001}

In protostellar cores and Bok globules the dispersion of the 
polarization angle is sometimes very large, and the Chandrasekhar--Fermi
formula is not expected to hold. However, the formula is sometimes applied 
irrespective of the large value of $\sigma_{\chi}$ (eg Itoh et al. 1999;
Davis et al. 2000; Henning et al. 2001). This is perhaps
justified by the expectation that the degree of topological complexity of
the magnetic field, and so the value of $\sigma_{\chi}$, should be related 
in some way to the rms Alfv\'{e}nic Mach number ${\cal M}_a$. 
A full numerical study of the dependence of $\sigma_{\chi}$ on ${\cal M}_a$
can be pursued with the MHD turbulence simulations, and will be presented
elsewhere. For the purpose of this work we test the Chandrasekhar--Fermi
formula on the cores C1, C2 and C3, and provide the values of the 
corresponding correction factor $f$.

\nocite{Itoh+99}

The turbulent velocity dispersion in protostellar 
cores and Bok globules is often estimated using the line width of
the CS molecule. It is possible that the bulk 
of the CS emission come from an envelope around the core, and not 
from the core interior, where most of the 
dust emission is generated, at least for the densest cores. This 
complicates the application of the Chandrasekhar--Fermi formula and its
interpretation. For this reason, we have tested the formula by defining
the turbulent velocity as given by the CS spectra. 

Maps of synthetic spectra of the J=1--0 transition of CS are computed using 
a non--LTE Monte Carlo radiative transfer code (Juvela 1997, 1998),
from the three dimensional density and velocity fields generated in
the numerical MHD experiment. The method of computing synthetic spectra 
was presented in Padoan et al. (1998). We assume a linear 
size of the computational box $L=6.25$~pc, a uniform temperature $T=10$~K, 
an rms turbulent velocity $\sigma_v=3.0$~km/s and an average gas density 
$\langle n \rangle=320$~cm$^{-3}$. These values are used to scale the 
numerical variables into physical units. In the radiative 
transfer computation the full three dimensional density 
and velocity fields are used. 
The radiative transfer is a non--local problem, and so it is solved
for the whole three dimensional computational box. 

\nocite{Juvela97}  \nocite{Juvela98}  \nocite{Padoan+98cat}

\ifnum\inlinefig=1
\begin{table}[!th]
\begin{tabular}{l|ccc|ccc|ccc}
\hline
\hline
Core   & \multicolumn{3}{|c|}{C1} & \multicolumn{3}{|c|}{C2}   & \multicolumn{3}{|c}{C3}  \\
                            & x    & y    & z    & x    & y    & z    & x    & y    & z    \\ 
\hline
$\sigma_v \,\,\, [km/s]$    & 0.96 & 0.96 & 0.82 & 0.94 & 1.34 & 1.04 & 1.00 & 0.73 & 1.32 \\
$\sigma_{\chi}$ \,\, [rad]  & 0.48 & 0.73 & 0.37 & 0.59 & 0.58 & 0.71 & 0.55 & 0.73 & 0.65 \\
          & (0.49) & (0.70) & (0.37) & (0.64) & (0.62) & (0.71) & (0.52) & (0.74) & (0.65) \\
$f$                         & 0.32 & 0.48 & 0.29 & 0.43 & 0.29 & 0.46 & 0.35 & 0.63 & 0.31 \\
          & (0.32) & (0.46) & (0.29) & (0.46) & (0.31) & (0.46) & (0.33) & (0.64) & (0.31) \\
\hline
\end{tabular}
\caption{%
Estimated parameters for the three cores C1, C2 and C3. From top to bottom: rms
turbulent velocity from CS spectral maps; rms polarization angle; correction 
factor for the Chandrasekhar--Fermi formula (see text). Values in parenthesis 
are for the case of unaligned grains 
above $A_{V,0}=3$~mag.}
\end{table}
\fi

Spectral maps are obtained for the same lines of sight that define the
polarization maps. The CS spectral maps that match the polarization maps
of Figure~5 are plotted in Figure~9. 
All the spectra are used to compute the rms velocity $\sigma_v$. 
This is defined as the rms velocity averaged
along the spectral profile, without any Gaussian fit:
\begin{equation}
\sigma_v =  \left(\frac{\int (v-\langle v\rangle)^2 \, T(v) \,\,dv}{\int T(v) \,\, dv}\right)^{1/2} 
\label{sigmav}
\end{equation}
where $T(v)$ is the antenna temperature at the velocity channel $v$, and 
the result is averaged over the whole spectral map of each core. 
The average velocity $\langle v\rangle$ is defined as
\begin{equation}
\langle v\rangle = \frac{\int v \, T(v) \,\,dv}{\int T(v) \,\, dv} 
\label{v}
\end{equation}
The values of $\sigma_v$ for three orthogonal lines of sight are 
listed in Table~1. The average value is $\sigma_v=1.01 \pm 0.19$~km/s.

In order to determine the value of the coefficient $f$, we finally
need to compute the rms Alfv\'{e}n velocity in the cores. This is
done directly in the three dimensional MHD data cube. 
Using the values of $\sigma_{\chi}$ computed in the previous section,
and the values of $\sigma_v$ and rms $v_a$ computed here, 
we can now evaluate the coefficient $f$. Results are listed in Table~1.
The average value of $f$ is found to be $f=0.40\pm0.11$ for both
models $A$ and $B$.

\section{Discussion}

In order to describe the formation of protostellar cores
we have simulated the dynamics of molecular cloud turbulence on a relatively
large scale ($\approx 6$~pc). Simulating a large scale is necessary because 
the boundary and initial 
conditions of a numerical simulation (and much more of an analytic model) are always
idealized in some way. For example, a numerical simulation or an analytic  
model describing the evolution and collapse of a single protostellar core
can hardly be considered as models for the core {\it formation}, since 
the core is obviously formed by the ad hoc setup of the initial conditions.
On the other hand, if a larger scale is simulated, the process of core 
formation is controlled by the gas dynamics on the large scale, and may even be 
rather insensitive to the specific initial and boundary conditions at the
large scale. 

In the simulation used for this work the typical size of cores is 
less than one tenth of the size of the computational box, and so we 
argue that our periodic boundary conditions do not affect too much 
the formation and evolution of dense cores. The setup of the initial 
conditions is also unlikely to affect dramatically the outcome of the 
simulation, since after one dynamical time the turbulent
flow begins to lose memory of its initial conditions. The random external
driving force can instead have important effects, in the sense that 
the flow would develop much more small scale structure if the forcing
was done on a small scale, and the number of correlation lengths of the
magnetic field, $N_{corr}$, could be relatively large. However, that 
would imply a power spectrum of turbulence peaking on small scales. 
In contrast, all observational evidence for the power spectrum of 
turbulence in molecular clouds is consistent with a scale free dynamics 
(Miesch \& Bally 1994; Ossenkopf \& Mac Low 2000), 
with most energy on the largest scale. The large scale 
driving used in the present simulation, is therefore consistent
with the observations. As a consequence, the number of correlation
lengths of the magnetic field must be relatively small, $N_{corr}\sim 1$.
This explains the relatively large values of $P$ found from our 
super--Alfv\'{e}nic simulations, since a significant cancellation
effect requires both a weak magnetic field (super--Alfv\'{e}nic
turbulence) and a large number of correlation lengths.

\nocite{Miesch+Bally94} \nocite{Ossenkopf+MacLow2000}

It is well established that the polarized extinction of background
stars at optical and near--IR wavelengths does not reveal the magnetic 
field direction in the dense interiors of molecular clouds, beyond an 
optical depth $A_v=1-2$~mag (Goodman et al. 1995; Arce et al. 1998). This is
generally interpreted as the effect of unaligned grains at large
visual extinction, since it can be shown that most alignment mechanisms
are ineffective at large optical depth (Lazarian, Goodman \& Myers 1997).
Wiebe \& Watson (2001), on the contrary, suggest that the
results of polarization studies of background stars can be
explained by a cancellation effect due to the magnetic field
tangling along the line of sight. This requires super--Alfv\'{e}nic 
turbulence and a large number of correlation lengths of the magnetic field, 
$N_{corr}$. They find that $N_{corr}\ge 10$ is necessary to explain 
the observations, and use a numerical MHD simulation by Stone, 
Ostriker \& Gammie (1998) as an example of a theoretical model
with $N_{corr}>10$. However, as explained in the previous paragraph,
there is no observational evidence in favor of a power spectrum
of turbulence with a peak at small scale, which is needed to 
produce a large value of $N_{corr}$. The simulation by Stone, 
Ostriker \& Gammie (1998) used by Wiebe and Watson (2001) 
is performed with a random driving on an intermediate scale,
which explains the large value of $N_{corr}$, but is certainly 
not motivated by the observations. 

\nocite{Wiebe+Watson2001}

In \S~5 we have seen that recent sub-mm polarization maps of 
quiescent cores always produce $P$--$I$ plots characterized
by an upper envelope with $P$ decreasing with increasing $I$.
This result is predicted by our model $B$, that is by assuming
that grains at visual extinction larger than $A_{V,0}=3$~mag are
not aligned. It is not reproduced by model $A$, where the grain
alignment is assumed to be independent of $A_{V}$, because
the magnetic field tangling in the cores is insufficient to
produce such drop in $P$. Although it is still possible
that higher resolution simulations following the core collapse
to higher densities generate a larger cancellation effect,
that remains to be proven. From the present calculations 
it seems that the reasonable assumptions of model B provide
a good interpretation of the observational $P$--$I$ plots.
If this interpretation is correct, we can conclude that 
sub-mm polarization maps
cannot probe the magnetic field structure inside quiescent cores,
at visual extinction larger than $A_{V,0}=3$~mag.
The use of such maps to constrain 
models of protostellar core formation and evolution is therefore
questionable, especially if the effect of unaligned grains is not
taken carefully into account. 
Polarization studies of sub-mm dust continuum emission from 
quiescent cores are therefore not inconsistent
with previous results of polarized extinction of background
stars at optical and near--IR wavelengths mentioned above.

The specific value of $A_{V,0}=3$~mag used to compute model $B$
is a ``conservative'' choice. A value of $A_{V,0}\approx 2$~mag,
or even lower, is not ruled out by our study. It is likely that 
the alignment efficiency of grains begins to decrease significantly
at $A_V=1-2$~mag, as suggested by theoretical studies (Lazarian, 
Goodman \& Myers 1997). In this work we show that at values
of visual extinction larger then 3~mag grains must be almost
completely unaligned. 

The MHD simulation used in this work has a limited numerical
resolution. A $128^3$ numerical mesh has been used and the 
volume selected around each core contains only $20^3$ computational 
cells. Differences in the polarization maps of the cores between 
models $A$ and $B$ (see Figure~5) are equally limited. A much higher 
numerical resolution could in principle generates even larger
differences between the magnetic field orientation in the
densest regions of the cores and in their lower density
envelopes. The center of the core at $A_V>3$~mag, for example, 
could have collapsed dragging in the magnetic field lines, 
and this process would be totally invisible in the polarization maps.
However, models for the formation and evolution of protostellar cores 
could be tested with polarization maps of quiescent cores if the comparison
were limited to the regions of low visual extinction around the cores. 

\nocite{Lazarian+97} 

We have shown in the previous section that the Chandrasekhar--Fermi
formula predicts the value of the rms Alfv\'{e}n speed in the cores, 
if a correction factor $f\approx 0.4$ is used, and if the rms turbulent 
velocity is derived from the line width of the J=1--0 CS transition. 
The value of $\sigma_{\chi}$ should be appropriately minimized as discussed
in the previous section. However, in order to infer a magnetic field strength
from an rms Alfv\'{e}n speed and $\sigma_{\chi}$, the gas density in the cores must 
be estimated, which introduces a significant uncertainty.

\section{Conclusions}

In this work we have computed dust continuum polarization from protostellar
cores, assembled by super--sonic turbulent flows in models of molecular clouds,
using numerical simulations of highly super--sonic MHD turbulence.
The results are compared with recent polarization maps of
quiescent protostellar cores and Bok globules.

The main results of this work are: i) Values of $P$ between 1 and 10\% (up to
almost $P_{max}$) are produced by super--Alfv\'{e}nic turbulence; 
ii) A steep decrease of $P$ with increasing $I$, as observed in a number 
of protostellar cores and Bok globules, is always found in self--gravitating
cores selected from the MHD simulation, if grains are not aligned above a 
certain value of visual extinction $A_{V,0}$ (model $B$); 
iii) The same behavior is hard to reproduce if
dust grains are aligned independently of $A_{V}$ (model $A$); iv) The 
Chandrasekhar--Fermi formula, corrected by a factor $f\approx 0.4$,
can be used to estimate the average magnetic field strength in the cores. 

We conclude that sub--mm polarization maps of quiescent cores are not 
likely to faithfully map
the magnetic field inside the cores at visual extinction larger than
$A_{V,0}\approx 3$~mag. The use of such maps to constrain models of 
protostellar core formation and evolution is questionable, especially
if the effect of unaligned grains is not properly taken into account. 
This conclusion shows that there is no inconsistency between the results
from optical and near--IR polarized absorption of background stars,
and the observed polarization of sub-mm dust continuum from
quiescent cores. In both cases, grains at large visual extinction
appear to be virtually unaligned.

\acknowledgements

This work was supported by NSF grant AST-9721455.
\AA ke Nordlund acknowledges partial support by the Danish National 
Research Foundation through its establishment of the Theoretical 
Astrophysics Center. 
Mika Juvela acknowledges support by the Academy of Finland 
Grant no. 1011055.

%\bibliographystyle{aabib}
%\bibliography{padoan,MC,GALAXIES}

\ifnum\inlinefig=0
\clearpage

\onecolumn

{\bf Figure captions:} 

{\bf Table \ref{tab1}:} Estimated parameters for the three cores C1, 
C2 and C3. From top to bottom: rms turbulent velocity from CS spectral 
maps; rms polarization angle; correction factor for the 
Chandrasekhar--Fermi formula (see text). Values in parenthesis 
are for the case of unaligned grains above $A_{V,0}=3$~mag.\\

{\bf Figure \ref{fig1}:} Grain cross section ratio as a function of 
grain axial ratio for prolate spheroids (dashed line) and oblate 
spheroids (solid line), at wavelength $\lambda=350$~$\mu$m. The 
grains are the ``astronomical silicate'' discussed by Draine \& Lee 
(1984), as modified by Li \& Draine (2001).\\

{\bf Figure \ref{fig2}:} Polarization map from the MHD model (model $A$). 
The average magnetic field is oriented in the vertical direction. The 
length of the polarization vectors is proportional to the degree of 
polarization, with the longest vector corresponding to $P=14$~\%. 
Only one polarization vector every three computational cells is plotted. 
The contour map shows the sub--mm dust continuum intensity $I$.
The position of the self--gravitating cores C1, C2 and C3 (see text in \S4) 
is also shown.\\

{\bf Figure \ref{fig3}:} As in Figure~1, but assuming unaligned grains above 
$A_{V,0}=3$~mag (model $B$).\\

{\bf Figure \ref{fig4}:} Left panel: Scatter plot of the degree of polarization 
versus the emission intensity from the entire map in Figure~2. Only the 
polarization vectors shown in Figure~2 are used, that is one position every 
three computational cells. The average degree of polarization is 
$\langle P\rangle=5.6$~\%. Right panel: As left panel, but from the entire 
map in Figure~3, that is with unaligned grains above $A_{V,0}=3$~mag.\\

{\bf Figure \ref{fig5}:} Upper panels: Polarization maps as in Figure~2, 
but limited to the regions around the cores C1, C2 and C3 (see text). 
The polarization vectors are here plotted for one position every two 
computational cells, and the largest polarization vectors (found in 
the core C1) corresponds to $P\approx 11.5$~\%. Lower panels: As upper 
panels, but from the map in Figure~3, that is with unaligned grains 
above $A_{V,0}=3$~mag. \\

{\bf Figure \ref{fig6}:} Left panels (model $A$): Scatter plots of the 
degree of polarization versus the emission intensity from the maps in 
Figure~5 (upper panels). Only the polarization vectors shown in Figure~5 
are used, that is one position every two computational cells. Right panels 
(model $B$): same as left panels, but from the maps in the lower panels of 
Figure~5, that is assuming unaligned grains above $A_{V,0}=3$~mag. \\

{\bf Figure \ref{fig7}:} Left panel (model $A$): Scatter plot of the degree 
of polarization versus the emission intensity for the core C1, seen from 
three orthogonal points of view. Right panels (model $B$): same as left panels, 
but with unaligned grains above $A_{V,0}=3$~mag. \\

{\bf Figure \ref{fig8}:} Left panels (model $A$): Histograms of the 
polarization angle $\chi$ from the polarization maps shown in Figure~5 
(upper panels). All the available polarization vectors have been used, 
that is one per computational cell. Right panels (model $B$): same 
as left panels, but from the polarization maps in the lower panels 
of Figure~5, that is with unaligned grains above $A_{V,0}=3$~mag. \\

{\bf Figure \ref{fig9}:} Spectral maps of the J=1--0 transition of CS. 
The maps are located around the cores C1, C2 and C3, from top to bottom 
respectively. The regions shown in the maps correspond to the regions 
included in the maps of Figure~5. The velocity axis ranges from -2.76 
to 2.76 km/s, and the antenna temperature axis from 0 to 3 K. \\

\clearpage
\begin{table}
\begin{tabular}{lccccccccc}
\hline
\hline
Core & $\sigma_v$ [km/s] & $\langle\rho\rangle$  [g/cm$^3$] & $\sigma_{\chi}$ [rad] & $B_{CF}$ [$\mu$G] & $B_{eq}$ & $\langle B_{sky}\rangle$ & $\langle B\rangle$ & $B_{CF}/\langle B\rangle=f$ & $B_{eq}/\langle B\rangle$\\ 
\hline
 C1  &  0.68   &  3.7e-20   &   0.73 (0.61)  & 63.4 (75.9) & 46.3  &  21.2  &  23.1  & 2.74 (3.29)  & 2.00 \\ 
 C2  &  0.67   &  3.3e-20   &   0.58 (0.48)  & 73.7 (89.0) & 42.9  &  21.3  &  28.9  & 2.55 (3.08)  & 1.48 \\ 
 C3  &  0.72   &  1.6e-20   &   0.73 (0.71)  & 43.2 (44.4) & 31.7  &  5.6   &  14.0  & 3.09 (3.17)  & 2.26 \\ 
\hline
\end{tabular}
\caption{}
\label{tab1}
\end{table}

\clearpage
\begin{figure}[!th]
%\centerline
{\epsfxsize=12.cm \epsfbox{draine.eps}}
\caption[]{}
\label{fig1}
\end{figure}

\clearpage
\begin{figure}[!th]
\centerline
{\epsfxsize=16cm \epsfbox{pol_map_c_dir2.eps}}
\caption[]{}
\label{fig2}
\end{figure}

\clearpage
\begin{figure}[!th]
\centerline
{\epsfxsize=16cm \epsfbox{pol_map_c_dir2_av3.eps}}
\caption[]{}
\label{fig3}
\end{figure}

\clearpage
\begin{figure}[!th]
{\epsfxsize=8cm \epsfbox{pol_int_c_dir2.eps}}
{\epsfxsize=8cm \epsfbox{pol_int_c_dir2_av3.eps}}
\caption[]{}
\label{fig4}
\end{figure}

\clearpage
\begin{figure}[!th]
{\epsfxsize=5.45cm \epsfbox{pol_map_c1.eps}}
{\epsfxsize=5.45cm \epsfbox{pol_map_c2.eps}}
{\epsfxsize=5.45cm \epsfbox{pol_map_c3.eps}}
{\epsfxsize=5.45cm \epsfbox{pol_map_c1_av3.eps}}
{\epsfxsize=5.45cm \epsfbox{pol_map_c2_av3.eps}}
{\epsfxsize=5.45cm \epsfbox{pol_map_c3_av3.eps}}
\caption[]{}
\label{fig5}
\end{figure}

\clearpage
\begin{figure}[!th]
{\epsfxsize=8.2cm \epsfbox{pol_int_c1_dir2.eps}}
{\epsfxsize=8.2cm \epsfbox{pol_int_c1_dir2_av3.eps}}
{\epsfxsize=8.2cm \epsfbox{pol_int_c2_dir2.eps}}
{\epsfxsize=8.2cm \epsfbox{pol_int_c2_dir2_av3.eps}}
{\epsfxsize=8.2cm \epsfbox{pol_int_c3_dir2.eps}}
{\epsfxsize=8.2cm \epsfbox{pol_int_c3_dir2_av3.eps}}
\caption[]{}
\label{fig6}
\end{figure}

\clearpage
\begin{figure}[!th]
{\epsfxsize=8.2cm \epsfbox{pol_int_c1_dir1.eps}}
{\epsfxsize=8.2cm \epsfbox{pol_int_c1_dir1_av3.eps}}
{\epsfxsize=8.2cm \epsfbox{pol_int_c1_dir2.eps}}
{\epsfxsize=8.2cm \epsfbox{pol_int_c1_dir2_av3.eps}}
{\epsfxsize=8.2cm \epsfbox{pol_int_c1_dir3.eps}}
{\epsfxsize=8.2cm \epsfbox{pol_int_c1_dir3_av3.eps}}
\caption[]{}
\label{fig7}
\end{figure}

\clearpage
\begin{figure}[!th]
{\epsfxsize=8.2cm \epsfbox{hist_theta_c1.eps}}
{\epsfxsize=8.2cm \epsfbox{hist_theta_c1_av3.eps}}
{\epsfxsize=8.2cm \epsfbox{hist_theta_c2.eps}}
{\epsfxsize=8.2cm \epsfbox{hist_theta_c2_av3.eps}}
{\epsfxsize=8.2cm \epsfbox{hist_theta_c3.eps}}
{\epsfxsize=8.2cm \epsfbox{hist_theta_c3_av3.eps}}
\caption[]{}
\label{fig8}
\end{figure}

\clearpage
\begin{figure}[!th]
{\epsfxsize=5.4cm \epsfbox{map_cs_c1.eps}}
{\epsfxsize=5.4cm \epsfbox{map_cs_c2.eps}}
{\epsfxsize=5.4cm \epsfbox{map_cs_c3.eps}}
\caption[]{}
\label{fig9}
\end{figure}

\fi

\end{document}